\documentclass[useAMS]{gCOM2e}

\usepackage[vlined, ruled]{algorithm2e}
\usepackage{graphicx}
\usepackage{epsfig}
\usepackage{epstopdf}

\title{Task swapping networks in distributed systems}

\author{Dohan Kim$^{\ast}$\thanks{$^\ast$ Email: dkim@airesearch.kr}\\  {\em{A.I. Research Co., 2537-1 Kyungwon Plaza 201, Sinheung-dong, Sujeong-gu, Seongnam-si, Kyunggi-do, 461-811, S. Korea}}\\\vspace{6pt}
}

\begin{document}
\maketitle
\begin{abstract}
In this paper we propose task swapping networks for task reassignments by using task swappings in distributed systems. Some classes of task reassignments are achieved by using iterative local task swappings between software agents in distributed systems. We use group-theoretic methods to find a minimum-length sequence of adjacent task swappings needed from a source task assignment to a target task assignment in a task swapping network of several well-known topologies.

\end{abstract}


\renewenvironment{classcode}{%
        \par\ifkeywords\addvspace{-20pt}\else\addvspace{11pt}\fi%
  \keywordfont
  \noindent{\bf C.R. Categories{\rm{:}}\ }\ignorespaces
}

\begin{classcode}
C.2.1; G.2.3
\end{classcode}

\section{Introduction}
A distributed system is defined as a collection of independent computers or processors that are connected by an arbitrary interconnection network~\cite{Ramakrishnan1991, Tanenbaum1995}. The objective of a task assignment~\cite{Ramakrishnan1991,Shin1990} in a distributed system is to find an optimal or suboptimal assignment of tasks to processors (or agents), while satisfying temporal and spatial constraints imposed on the system. A task assignment is either preemptive or non-preemptive~\cite{Milojicic2000}. If a task assignment is preemptive, a task reassignment is allowed in such a way that  tasks are transferred between processors (or agents) during their execution for improving the system performance~\cite{Milojicic2000, Robinson1996}. Recent advances in software agent technology~\cite{Jennings1996, Kalinowski2010, Nwana1996} in distributed systems allow software entities to observe their environment, and cooperate with other entities if necessary to accomplish their goals. Task migration between software agents intends to improve the system throughput in a distributed system in which the loads incurred by tasks vary over time~\cite{Kalinowski2010}. A task reassignment can be achieved by iterative local task swappings, where a task swapping involves task migrations between a pair of agents as a method of local task reassignment~\cite{Chen2008}. A subclass of task assignment (or reassignment) problems involves an equal number of tasks and agents, finding a bijective task assignment between tasks and agents in such a way that the total task assignment (or reassignment) benefit is maximized~\cite{Zavlanos2008}.  Those bijective task assignment problems and their variants appear in a wide variety of areas in computer science and mathematics~\cite{Bowen1992, Burkard2009, Zavlanos2007, Zavlanos2008}. Given $n$ tasks and $n$ agents whose connections are described by a (connected) network topology, task swapping cost between two agents often relies on a distance in the network topology; the larger the distance between two agents in the network topology, the larger communication delays of migrating tasks caused by the network. Therefore, we need to consider how task swappings are performed on a given network topology for task reassignments. In this type of problems a group theory can be used to represent task reassignments including iterative task swappings. A group-theoretic approach to representing task assignments or reassignments have already been researched~\cite{Kim2010, Rowe2002}. However, little work has been done about task swappings for task reassignments between tasks and agents by using group theory. We propose a group-theoretic model of global task reassignments involving $n$ tasks and $n$ agents by using local task swappings in a distributed network of several well-known topologies. 

This paper is organized as follows. Section~\ref{sec:ProblemFormulation} presents the problem formulation and its assumptions. We give the necessary definitions for the problem formulation in this section. Section~\ref{sec:TranspositionGraphs} gives an introduction to permutation groups and Cayley graphs. We also discuss transposition graphs of several network topologies and their relationship to permutation sortings in this section. We present task swapping graphs and their examples in distributed systems in Section~\ref{sec:TaskSwappingNetworks}. Group-theoretic properties of task swapping graphs and their examples are discussed in this section. Section~\ref{sec:RelatedWork} summarizes the related work and implementation. Finally, we conclude in Section~\ref{sec:Conclusion}.

\section{Problem formulation and assumptions}
\label{sec:ProblemFormulation}

An \emph{agent graph} $G_a = (V_a, E_a)$~\cite{Enokido2005} is an undirected graph, where $V_a$ denotes a set of agents in a distributed system and $E_a$ denotes communication links between agents. Let $T = \{t_1, t_2,\ldots, t_n\}$ be a set of $n$ tasks and $A = \{a_1, a_2,\ldots, a_n\}$ be a set of $n$ agents represented by an agent graph $G_a = (V_a, E_a)$. Now, each task is assigned to each agent in $G_a = (V_a, E_a)$ bijectively as an initial task assignment by using a task assignment algorithm. Suppose the load on each agent varies over time, implying that an initial task assignment may not remain optimal or suboptimal. The \emph{load rebalancing} intends to find a task reassignment in order to decrease makespan~\cite{Aggarwal2003}, or to achieve a target load balance level along with minimum task migration cost~\cite{Chen2008}. Task migration\footnote{In this paper we use \emph{task migration} and \emph{process migration} interchangeably.} between agents incurs a task migration cost~\cite{Du2007} to transfer the state of the running task (e.g., execution state, I/O state, etc) of one agent to the other agent. This cost is not trivial, following that only nearby agents (or processors) in the network topology are often involved to perform a task migration~\cite{Chen2008,Hu1998}. Moreover, if we restrict each task reassignment to be bijective between tasks and agents, then iterative local task swappings can be used for a global task reassignment. A \emph{swapping distance} is defined as the distance between agents in an agent graph $G_a$ for a task swapping. For example, a task swapping of swapping distance 1 is a task swapping between adjacent agents in $G_a$, and a task swapping of swapping distance 2 is a task swapping between agents whose distance is 2 in $G_a$. Now, the overall cost  difference before task swapping and after task swapping for computationally intensive tasks (rather than communicationally intensive tasks) is roughly the cost of a target task assignment after task swapping subtracted by the cost of a source task assignment before task swapping, and added by the task swapping cost itself. In the remainder of this paper we assume the followings:
\begin{enumerate}
\item Each task and agent are not necessarily homogeneous, and the load on each agent may vary over time.
\item An agent network described by $G_a = (V_a, E_a)$ is static without taking mobility (e.g., dynamic join or leave by agents~\cite{Zhang2006}) into account.
\item A task swapping cost is uniform between adjacent agents in the agent network.
\item A task swapping cost is assumed to be proportional to a task swapping distance. 
\item Each task is long-lived and computationally intensive.
\item A global task reassignment is obtained by using iterative and sequential local task swappings.
\item A task startup cost and the message passing overhead for task migration protocol are ignored.
\item Every agent in $G_a = (V_a, E_a)$ is cooperative~\cite{Sycara1998}, pursuing the same goal to improve their collective performance for task assignments.
\end{enumerate}
It follows that the expected total cost of task migrations hinges on how many local task swappings are needed from a source task assignment to a target task assignment. A decision making by the coordinator agent(s) whether or not a task reassignment is performed is based on the information regarding the total cost of task migrations along with the cost difference involving a source and target task assignment. Let $g_1$ be a bijective source task assignment between a set of $n$ tasks $T = \{t_1, t_2,\ldots, t_n\}$ and a set of $n$ agents $A = \{a_1, a_2,\ldots, a_n\}$ before task migrations. Let $g_2$ be a feasible target task assignment between $T$ and $A$ after task migrations, which implies that $g_2$ can be obtained by using iterative local task swappings. Let $h_1$ be the cost of task assignment $g_1$ and let $h_2$ be the cost of task assignment $g_2$. Let $f(g_1, g_2, s_1, s_2,\ldots, s_k)$ be the total cost of task migrations, where $s_i$ for $1\leq i \leq k < n$ is the number of local task swappings of swapping distance $i$ involved in converting $g_1$ to $g_2$. Then, a task reassignment benefit $b(h_1, h_2, f)$ is defined as $-(h_2 - h_1 + f(g_1, g_2, s_1, s_2,\ldots, s_k))$, i.e., $h_1 - h_2 - f(g_1, g_2, s_1, s_2,\ldots, s_k)$. The higer value of $b(h_1, h_2, f)$ implies that a task reassignment is more desirable, while the negative value of $b(h_1, h_2, f)$ implies that a task reassignment from $g_1$ to $g_2$ is not desirable at all. If we restrict a local task swapping to be a task swapping between adjacent agents in $G_a = (V_a, E_a)$, $f(g_1, g_2, s_1, s_2,\ldots, s_k)$ is simply $cs_1$, where $c$ is constant. If we restrict a local task swapping to be a task swapping between agents in $G_a = (V_a, E_a)$ within distance $m$, $f(g_1, g_2, s_1, s_2,\ldots, s_k)$ is $cs_1 + 2cs_2 + \cdots + mcs_m$, where $c$ is constant and $s_r$ $(1 \leq r \leq m)$ is the number of task swappings of swapping distance $r$ involved in converting $g_1$ to $g_2$. The problem is formulated as follows:

Given a source task assignment $g_1$ and a feasible target task assignment $g_2$ on the agent network described by an agent graph $G_a = (V_a, E_a)$, find the minimum total cost of task migrations $f(g_1, g_2, s_1, s_2, \ldots, s_k)$ to reach from $g_1$ to $g_2$.

In this paper we only concern local task swappings of swapping distance 1, which are adjacent task swappings between agents in $G_a = (V_a, E_a)$. Since we assume that a task swapping cost is uniform between adjacent agents in $G_a = (V_a, E_a)$, the problem is reduced to find the minimum number of adjacent task swappings needed from a source task assignment to reach a target task assignment in $G_a = (V_a, E_a)$.

\section{Transposition and Cayley graphs}
\label{sec:TranspositionGraphs}
We first give a brief introduction to finite groups found in~\cite{Fraleigh1998, Hungerford1980, Heydemann1997, Kim2010}.

A \emph{group} $(G,\,\cdot\,)$ is a nonempty set \emph{G}, closed under a binary operation $\cdot:G \times G \rightarrow G$, such that the following axioms are satisfied:
(i) $(a\cdot b)\cdot c =  a \cdot (b \cdot c)$ for all $a,b,c \in G$; (ii) there is an element \emph{e} $\in$ \emph{G} such that for all $x\in G,~e \cdot x = x \cdot e = x$; (iii) for each element $a \in G$, there is an element $a^{-1} \in G$ such that $a \cdot a^{-1}= a^{-1} \cdot a = e$. If $H$ is a nonempty subset of $G$ and is also a group under the binary operation $\cdot$ in $G$, then $H$ is called a \emph{subgroup} of $G$. 

The group of all bijections $I_n  \rightarrow I_n$, where $I_n = \{1, 2,\ldots, n\}$, is called the \emph{symmetric group on n letters} and denoted $\mathfrak{S}_n$. Since $\mathfrak{S}_n$ is the group of all permutations of $I_n = \{1, 2,\ldots, n\}$, it has order $n!$. A \emph{permutation group} is a subgroup of some $\mathfrak{S}_n$.  

Let $i_1, i_2,\ldots, i_n$ be distinct elements of $I_n = \{1, 2,\ldots, n\}$. Then, \\
$\begin{pmatrix} 
1&2&\cdots&n\\
i_1&i_2 &\cdots&i_n
\end{pmatrix}$
$\stackrel{\rm{def}}{=}i_1\,i_2\,\cdots\,i_n$$ \in \mathfrak{S}_n$
denotes the permutation that maps $1 \mapsto i_1,2 \mapsto i_2, \ldots, n \mapsto i_n$. 

Let $i_1, i_2,\ldots, i_r\,(r \leq n)$ be distinct elements of $I
_n = \{1, 2,\ldots, n\}$. Then $(i_1\,i_2\cdots i_r)$ is defined as the permutation that maps $i_1\mapsto i_2$, $i_2\mapsto i_3$,\ldots, $i_{r-1}$$ \mapsto i_r$ and $i_r\mapsto i_1$, and every other element of $I_n$ maps onto itself.  Then, $(i_1\,i_2 \cdots i_r)$ is called a cycle of length $r$ or an $r$-cycle; a 2-cycle is called a \emph{transposition}.

Let $G$ be a group and let $s_i \in G$ for $i \in I$. The subgroup generated by $S=\{s_i : i \in I\}$ is the smallest subgroup of $G$ containing the set $S$. If this subgroup is all of $G$, then $S$ is called a \emph{generating set} of $G$.

An (right) action of a group $G$ on a set $X$ is a function $X \times G \rightarrow X$ (usually denoted by $(x,\,g) \mapsto xg$) such that for all $x \in X$ and $g_1, g_2 \in G$: (i) $xe = x$; (ii) $x(g_1g_2) = (xg_1)g_2$. When such an action is given, we say that $G$ acts (right) on the set $X$. The set $X$ is called a (right) $G$-set. If $X$ is $G$ as a set, we say that $G$ acts on itself.

\begin{theorem}[\cite{Hungerford1980}]
\label{theorem:permutationsofSn}
Every non-identity permutation in $\mathfrak{S}_n$ can be expressed as a product of disjoint cycles of length at least 2. Further, $\mathfrak{S}_n$ can be expressed as a product of (not necessarily disjoint) transpositions. $\qquad$  
\end{theorem}

For example, $p=2\,1\,6\,3\,4\,5 \in \mathfrak{S}_6$ is written as $(1\,2)(3\,6\,5\,4)$ as a product of disjoint cycles or $(1\,2)(3\,4)(3\,5)(3\,6)$ as a product of transpositions.

Let $G$ be a finite group and $S$ be a generating set of $G$.  Then, $\text{Cay}(G, S)$ denotes the \emph{Cayley graph}~\cite{Akers1989, Heydemann1997, Lakshmivarahan1993} of $G$ with the generating set $S$, where the set of vertices $V$ of $\text{Cay}(G, S)$ corresponds to the elements of $G$, and the set of edges $E$ of $\text{Cay}(G, S)$ corresponds to the action of generators in such a way that $E=\{<x, y>_g: x, y \in G \text{ and } g \in S \text{ such that } y=x\cdot g\}$. In this paper we assume that $G$ is a finite permutation group and the elements of $S$ are transpositions in $\text{Cay}(G, S)$. It follows that $S$ is closed under inverses, and $E$ is undirected. In other words, an edge $<x, y>$ of the resulting $\text{Cay}(G, S)$ is viewed as both $<x, y>_g$ and $<y, x>_g$.
If $G$ is a finite group and $S$ is a set of transpositions that generates $G$, then a \emph{transposition graph} $T = (<n>, S)$~\cite{ Lakshmivarahan1993} is an undirected graph in which $<n>$ denotes a vertex set of cardinality $n$ and each edge $<i,\,j>$ denotes transposition $(i\,j)$. If a transposition graph is a tree, we call the resulting transposition graph as a \emph{transposition tree}. We next provide examples of the generating sets for permutation groups including their transposition trees. Let $S_1=\{(i\,i+1):1 \leq i < n\},
S_2=\{(2i-1\,2i):1 \leq i \leq n\},
S_3=\{(1\,i):2 \leq i \leq n\},
S_4=\{(i\,j):1 \leq i <j \leq n\},
S_5=\{(i\,j):1 \leq i \leq k <j \leq n\},$ and $S_6=S_1 \cup \{(1\,n)\}$.

\begin{figure}[h!]
\centering
\includegraphics[width=0.99\textwidth]{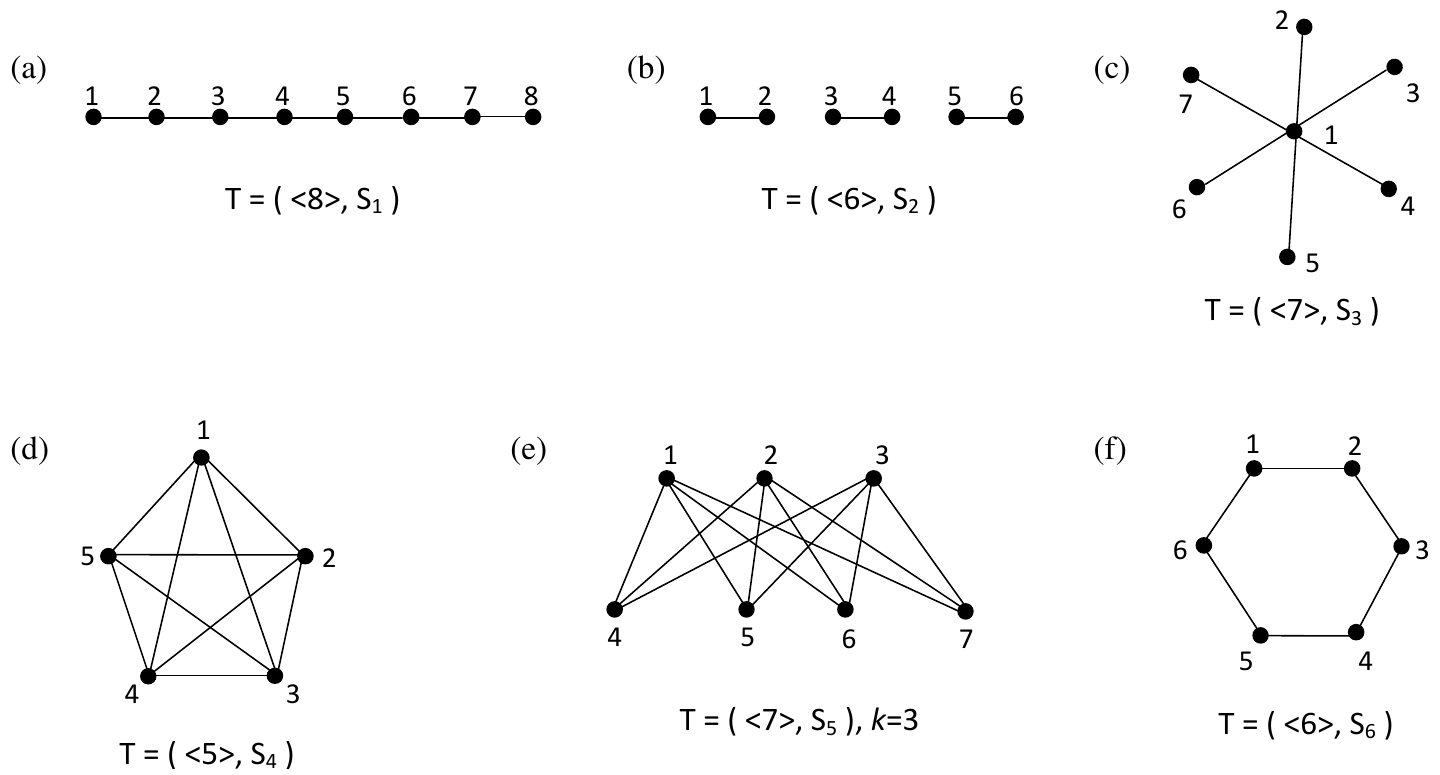}
\caption{Examples of transposition graphs $T=(<n>, S)$~\cite{ Lakshmivarahan1993}.}
\label{fig:TranspositionGraph}
\end{figure}

Figure~\ref{fig:TranspositionGraph} shows different kinds of transposition graphs corresponding to the generating sets $S_1,\,S_2,\ldots,\,S_6$ for some $n$ and $k$. In this paper a transposition graph having $n$ vertices are labeled from 1 to $n$ without any duplication. The Cayley graph generated by $S_1$ is called the \emph{bubble-sort} graph $\text{BS}_n$~\cite{Akers1989, Lakshmivarahan1993}, by $S_2$ is called the \emph{hypercube} graph $\text{HC}_n$~\cite{Heydemann1997}, by $S_3$ is called the \emph{star} graph $\text{ST}_n$~\cite{Akers1989, Lakshmivarahan1993,  Heydemann1997}, by $S_4$ is called the \emph{complete transposition} graph $\text{CT}_n$~\cite{Lakshmivarahan1993}, by $S_5$ is called the \emph{generalized star} graph $\text{GST}_{n, k}$~\cite{Heydemann1997}, and by $S_6$ is called the \emph{modified bubble sort} graph $\text{MBS}_n$~\cite{Lakshmivarahan1993}.

A path $p$ from vertex $v_1$ to vertex $v_2$ in $\text{Cay}(G, S)$ can be represented by a sequence of generators $g_1, g_2,\ldots, g_k$, where $g_i \in S$ for $1 \leq i \leq k$. By abuse of notation, we let $p=g_1 g_2\ldots g_k$. Note that $p$ is also a path from vertex $v_2^{-1}v_1$ to vertex $I$ (i.e., $v_1p=v_2$ if and only if $v_2^{-1}v_1p=I$ for the identity permutation $I$ in $G$). Thus, to find a path from vertex $v_1$ to vertex $v_2$ is reduced to find a path from vertex $v_2^{-1}v_1$ to vertex $I$, which in turn is reduced to the problem of sorting $v_2^{-1}v_1$ to $I$ using the generating set $S$~\cite{Akers1989}.

We now consider the bubble sort (Cayley) graph $\text{BS}_4$ corresponding to the transposition tree $T=(<4>, S_1)$. Since adjacent transpositions $(1\,2), (2\,3)$, and $(3\,4)$ generate $\mathfrak{S}_4$, $\text{BS}_4$ has $|\mathfrak{S}_4|=24$ vertices, each of which has degree 3. A shortest path from permutation $p_1$ to permutation $p_2$ in $\text{BS}_4$ is obtained by using the bubble sort algorithm~\cite{Knuth1973}. For example, let vertex $v_1$ be $2\,1\,3\,4$ and vertex $v_2$ be $3\,1\,4\,2$ in $\text{BS}_4$. Then, to find path $p$ from $v_1=2\,1\,3\,4$ to $v_2=3\,1\,4\,2$ is equivalent to find path $p$ from
$v_2^{-1}v_1=4\,2\,1\,3$ to permutation $I$: $4\,2\,1\,3 \xrightarrow{<1, 2>}$$ 2\,4\,1\,3 \xrightarrow{<2, 3>} $$2\,1\,4\,3 \xrightarrow{<3, 4>} $$2\,1\,3\,4 \xrightarrow{<1, 2>} $$1\,2\,3\,4$, 
where the label of each arrow denotes an edge in $\text{BS}_4$. Therefore, we have $p=(1\,2)$$(2\,3)$$(3\,4)$$(1\,2)$. 

As an another example, consider the star (Cayley) graph $\text{ ST}_4$ corresponding to the transposition tree $T=(<4>, S_3)$. Similarly to the bubble sort graph $\text{ BS}_4$, star transpositions $(1\,2), (1\,3)$, and $(1\,4)$ generate $\mathfrak{S}_4$. Therefore, $\text{ST}_4$ has $|\mathfrak{S}_4|=24$ vertices, each of which has degree 3. 

Note that transpositions denoted by a transposition tree of order $n$ labeled from 1 to $n$ generate $\mathfrak{S}_n$~\cite{Akers1989}. However, not every transposition graph having $n\,(n \geq 2)$ vertices yields a Cayley graph of $\mathfrak{S}_n$. For example, Figure~\ref{fig:TranspositionGraph}(b) shows a transposition graph having 6 vertices, but it yields a Cayley graph of a group isomorphic to $C_2 \times C_2 \times C_2$ instead of $\mathfrak{S}_6$, where $C_2$ is the cyclic group of order 2. 

A \emph{minimum generator sequence}~\cite{Jerrum1985} for permutation $p$ of a permutation group $G$ using the generating set $S$ is a minimum-length sequence consisting of generators in $S$ whose composition is $p$. For example, permutation $\pi=4\,2\,3\,1 \in \mathfrak{S}_4$ can be expressed as $\pi=(1\,2)(2\,3)(3\,4)(2\,3)(1\,2)$ of length 5 or $\pi=(1\,4)$ of length 1 using the generating set $S_6^\prime=\{(i\,i+1):1 \leq i < 4\} \cup \{(1\,4)\}$ in which $\pi=(1\,4)$ is the minimum generator sequence of length 1. Further, if an element $s$ in the generating set $S$ can be expressed as the product of elements in $S$ other than $s$, then the generating set is called \emph{redundant}~\cite{Lakshmivarahan1993}. For example, $S_6$ is redundant, since $(1\,n)$ can be expressed as $(1\,2)(2\,3)\cdots(n-1\,n)\cdots(2\,3)(1\,2)$. 

Recall that the \emph{diameter}~\cite{Enyioha2009} of a connected graph is the length of the "longest shortest path" between two vertices of the graph. Thus, the diameter of $\text{Cay}(G, S)$ is an upper bound of distance $d(\sigma, I)$ from an arbitrary vertex $\sigma$ to vertex $I$ in $\text{Cay}(G, S)$, where the computation of $d(\sigma, I)$ is obtained by sorting permutation $\sigma$ to the identity permutation $I$ in $G$ by means of the minimum generator sequence using the generating set $S$~\cite{Heydemann1997}. Thus, the diameter of $\text{Cay}(G, S)$ is an upper bound of the lengths of minimum generator sequences for permutations in $G$ using the generating set $S$~\cite{Akers1989}.

\begin{table}[h!]
\centering
\caption{Properties of some known Cayley graphs~\cite{Lakshmivarahan1993}.}
\begin{tabular}{c|c|c|c}

\small $\text{Types}$ & \small $\text{Number of Vertices}$ & \small $\text{Degree}$ & \small $\text{Diameter}$\\
\hline
\small $\text{BS}_n$ & \small $n!$ & \small $n-1$ & \small $n(n-1)/2$   \\
%\hline
\small $\text{ST}_n$ &  \small $n!$ & \small $n-1$ & \small $\lfloor3(n-1)/2\rfloor$     \\
%\hline
\small $\text{CT}_n$ &  \small $n!$ & \small $n(n-1)/2$ & \small $n-1$     \\
%\hline
\small $\text{GST}_{n,k}$ & \small $n!$ & \small $k(n-k)$ & \small $n-1+ \max{(\lfloor k/2  \rfloor,  \lfloor (n-k)/2 \rfloor)} $  \\
\small $\text{MBS}_n$ & \small $n!$ & \small $n$ & \small $\text{Unknown}$     \\
\small $\text{HC}_{n}$ & \small $2^n$ & \small $n$ & \small $n$  \\

\end{tabular}
\label{table:DiameterOfCayleyGraphs}
\end{table}

Now, consider a \emph{permutation puzzle}~\cite{Akers1989} on a transposition tree described as follows: Given a transposition tree $T$ having $n$ vertices labeled from 1 to $n$, place $n$ markers, each of which is labeled from 1 to $n$, at the vertices of $T$ arbitrarily in such a way that each vertex of $T$ is paired with an exactly one marker. Let $P$ be such an arbitrary initial position of markers. Each legal move of the puzzle is to interchange the markers placed at the ends of an edge in $T$. The terminal position of markers, denoted $Q$, is the position in which each marker is paired with each vertex of $T$ having the same label. The puzzle is to find a sequence of legal moves  from a given initial position $P$ to the final position $Q$ with the minimum number of legal moves.

\begin{theorem}[\cite{Akers1989}]
\label{theorem:TranspositionTree}
Let $T$ be a transposition tree having $n$ vertices. Given an initial position $P$ for a permutation puzzle on $T$, the final position $Q$ for the puzzle can be reached by legal moves in the following number of steps
\begin{center}
$c(p) -n + \displaystyle\sum_{i=1}^n{d(i, p(i))}$, 
\end{center}
where $p$ is the permutation for $P$ as an assignment of markers to the vertices of $T$, $c(p)$ is the number of cycles in $p$, and $d(i, j)$ is the distance between vertex $i$ and vertex $j$ in $T$. $\qquad$
\end{theorem}

\begin{corollary}[\cite{Heydemann1997}]
\label{cor:diameterOfCayleygraph}
Let $\text{Diam(Cay(G, S))}$ be the diameter of $\text{Cay(G, S)}$ of a transposition tree $T$ having $n$ vertices. Let $P$ be an initial position and $Q$ be the final position for a permutation puzzle on $T$. Then,
\begin{center}
$\text{Diam(Cay(G, S))} \leq \displaystyle\max_{p \in G}\{c(p) -n + \sum_{i=1}^n{d(i, p(i))}\}$, 
\end{center}
where $p$ is the permutation for $P$ as an assignment of markers to the vertices of $T$, $c(p)$ is the number of cycles in $p$, and $d(i, j)$ is the distance between vertex $i$ and vertex $j$ in $T$. $\qquad$
\end{corollary}

The $\text{Cay(G, S)}$ of a transposition tree $T$ can be viewed as the state diagram of a permutation puzzle on $T$, where the vertices of $\text{Cay(G, S)}$ is the possible positions of markers on $T$ and each edge of $\text{Cay(G, S)}$ corresponds to a legal move of the permutation puzzle on $T$. An upper bound of the minimum number of legal moves to reach the final position $Q$ for a given initial position $P$ is indeed the diameter of  $\text{Cay(G, S)}$~\cite{Akers1989} of a transposition tree $T$. Moreover, the inequality in Corollary~\ref{cor:diameterOfCayleygraph} can be replaced by the equality for the cases of bubble sort and star graphs~\cite{Heydemann1997}. 

We close this section by describing the main ideas behind Theorem~\ref{theorem:TranspositionTree} (an interested reader may refer to~\cite{Akers1989} for further details). Consider a permutation puzzle on a transposition tree $T$ having $n$ vertices. Let $p$ be a permutation for position $P$ as an assignment of $n$ markers to the $n$ vertices of $T$. The marker $i$ is said to be \emph{homed} if $i=p(i)$. If there exists any unhomed marker in the assignment, either one of two cases occurs. The first case is that there exists an edge of $T$ involving two unhomed markers $x$ and $y$ such that they need to move toward each other for the final position $Q$. The second case is that there exists an edge of $T$ involving one unhomed marker $x$ and one homed marker $y$ such that $x$ needs to move toward $y$ for the final position $Q$. In both cases, interchanging markers $x$ and $y$ reduces the number of steps in Theorem~\ref{theorem:TranspositionTree}. Finally, if there is no such case in the assignment (i.e., all markers are homed), then the number is reduced to 0.

\section{Task swapping networks}
\label{sec:TaskSwappingNetworks}

In Section~\ref{sec:TranspositionGraphs} we discussed a permutation puzzle on a transposition tree $T$ having $n$ vertices and its legal moves. We now consider a transposition tree as a network topology in which each vertex of $T$ corresponds to an agent and each marker to a task. Further, consider a position of $n$ markers placed at $n$ vertices of $T$ as a task assignment involving $n$ tasks and $n$ agents. Each legal move of a permutation puzzle then corresponds to an adjacent task swapping in the network topology. We apply the idea of the permutation puzzle on a transposition tree to task assignments in a task swapping network. We first define our \emph{task swapping graph}.

\begin{figure}[h!]
\centering
\includegraphics[width=0.99\textwidth]{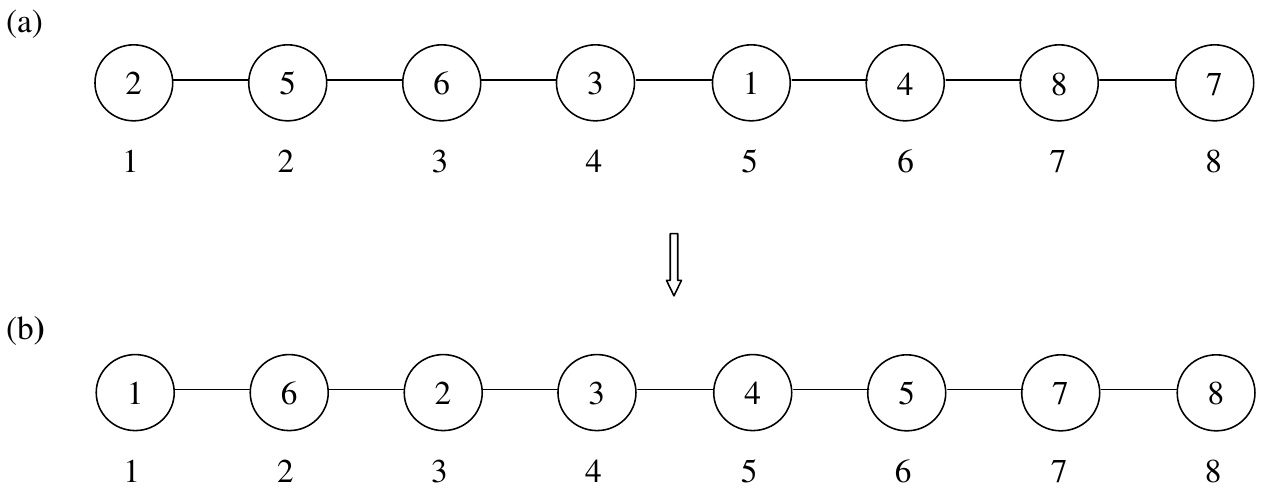}
\caption{Task swapping graphs of line topology.}
\label{fig:LineTopology}
\end{figure}

\begin{figure}[h!]
\centering
\includegraphics[width=0.99\textwidth]{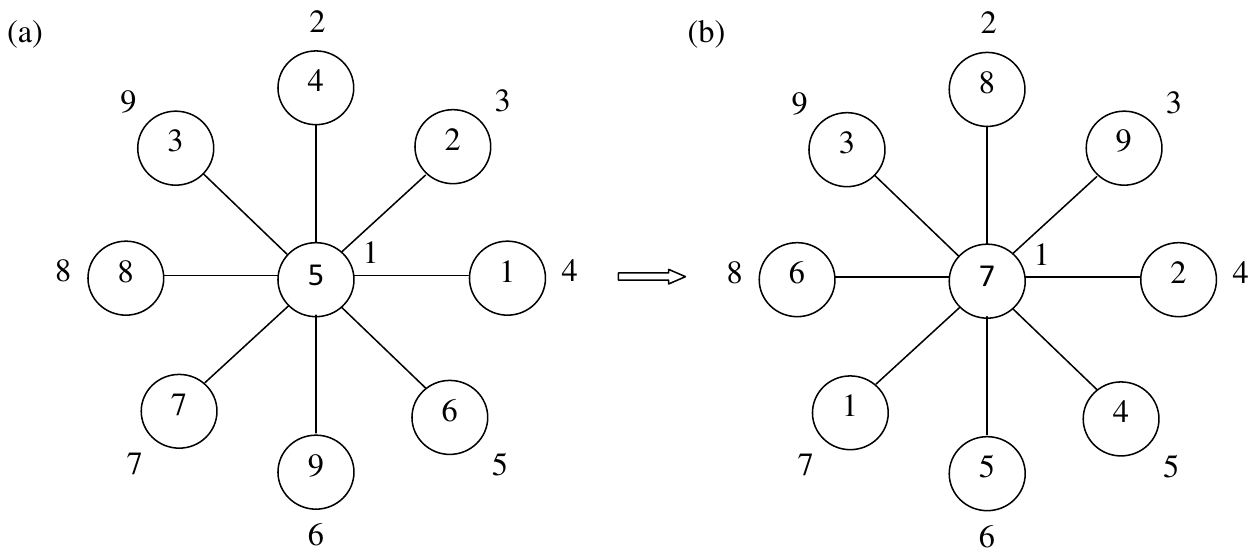}
\caption{Task swapping graphs of star topology.}
\label{fig:StarTopology}
\end{figure}

A \emph{task swapping graph} $\Gamma$ having $n$ vertices is a connected graph in which each vertex represents an agent and each edge represents a direct communication between agents. The label next to each vertex denotes an agent ID in $\Gamma$. Now, $n$ tasks are assigned to $n$ vertices of $\Gamma$ as a task assignment in which the label of each vertex in $\Gamma$ denotes a task ID. We assume that each task and agent ID are distinct numbers from the set $\{1, 2,\ldots, n\}$, where $n$ is the number of vertices in $\Gamma$. Therefore, each task assignment in $\Gamma$ is represented by permutation $p=\begin{pmatrix} 
1&2&\cdots&n	\\
t_1&t_2 &\cdots&t_n
\end{pmatrix}$
$ \in \mathfrak{S}_n$, where each element in the first row of $p$ denotes an agent ID and each element in the second row of $p$ denotes a task ID. If it is clear from the context, we simply denote $p$ as a one-line notation $t_1\,t_2\,\cdots\,t_n \in \mathfrak{S}_n$. For example, task assignments in the task swapping graphs in Figure~\ref{fig:LineTopology}(a) and Figure~\ref{fig:StarTopology}(a) are represented by $2\,5\,6\,3\,1\,4\,8\,7$ and $5\,4\,2\,1\,6\,9\,7\,8\,3$, respectively. Let $p \in \mathfrak{S}_n$ be a permutation representing a task assignment in $\Gamma$. Then, we denote this labeled task swapping graph $\Gamma$ as $\Gamma_p$. Now, we define a task swapping of swapping distance $k$ in $\Gamma_p$. A task swapping of swapping distance $k$ is the swapping of tasks between agents whose distance is $k$ in $\Gamma_p$. Recall that a right multiplication of permutation $p$ by transposition $(a\,b)$ exchanges the values in position $a$ and position $b$ of $p$. It follows that a task swapping of swapping distance $k$ in $\Gamma_p$ is represented by a right multiplication of $p$ by transposition $t=(i(v_1)\,i(v_2)) \in \mathfrak{S}_n$, where the distance between vertex $v_1$ and vertex $v_2$ is $k$ and $i(v)$ denotes the agent ID for vertex $v$ in $\Gamma_p$. Therefore, it converts $\Gamma_p$ into $\Gamma_{pt}$.

A \emph{task swapping network} is simply a distributed network represented by a task swapping graph $\Gamma$. In this paper a task swapping network is referred to as a task swapping graph $\Gamma$ unless otherwise stated. Based on the assumptions and the problem formulation in Section~\ref{sec:ProblemFormulation}, the problem is now reduced and rephrased by using a task swapping graph $\Gamma$:

Given a source task assignment $t_1$ and a feasible target task assignment $t_2$ in a task swapping graph $\Gamma$, find a minimum-length sequence of adjacent task swappings (i.e., task swappings of swapping distance 1) needed from $\Gamma_{t_1}$ to reach $\Gamma_{t_2}$. If $t_1 = t_2$, then nothing needs to be done. Similarly, a task swapping graph having one vertex involves no task swapping, which returns the empty sequence. A task swapping graph of two vertices involves only a single task swapping, which gives an immediate solution. In the remainder of this paper we assume that $t_1 \neq t_2$ and a task swapping graph has at least three vertices unless otherwise stated.

We find the solution of the problem for task swapping graphs of several key topologies. We first find the solution of the problem for a task swapping graph of line topology.

Line topology~\cite{Zhang2006, Gong2004, Gong2005, Zavlanos2008} is one of the simplest interconnection network topologies, where each agent is connected to exactly two neighboring agents other than the two end agents that are connected to only one neighboring agent (see Figure~\ref{fig:LineTopology}). We assume that each agent in a task swapping graph of line topology is labeled in ascending order from left to right as shown in Figure~\ref{fig:LineTopology}.  A task swapping graph of line topology having $n$ agents with their task assignment represented by permutation $p \in \mathfrak{S}_n$ is denoted as $\Gamma_p^{L(n)}$. For example, a task swapping graph of Figure~\ref{fig:LineTopology}(a) is denoted as $\Gamma_{p_1}^{L(8)}$ and a task swapping graph of Figure~\ref{fig:LineTopology}(b) is denoted as $\Gamma_{p_2}^{L(8)}$, respectively,  where $p_1=2\,5\,6\,3\,1\,4\,8\,7 \in \mathfrak{S}_8$ and $p_2=1\,6\,2\,3\,4\,5\,7\,8 \in \mathfrak{S}_8$. Recall that a right multiplication of permutation $p$ by transposition $(i\,i+1)$ exchanges the values in position $i$ and position $i+1$ of $p$. Since only adjacent task swappings are allowed, we use the generating set $S_1=\{(i\,i+1) : 1 \leq i < n\}$ to find a minimum-length sequence of adjacent task swappings needed from $\Gamma_{\pi_1}^{L(n)}$ for $\pi_1 \in \mathfrak{S}_n$ to reach $\Gamma_{\pi_2}^{L(n)}$ for $\pi_2 \in \mathfrak{S}_n$. For example, the number of the minimum adjacent task swappings needed from $\Gamma_{p_1}^{L(8)}$ in Figure~\ref {fig:LineTopology}(a) to reach $\Gamma_{p_2}^{L(8)}$ in Figure~\ref {fig:LineTopology}(b) is the minimum length of a permutation factorization of ${p_1}^{-1}p_2$ using the generating set $S_1^\prime=\{(i\,i+1) : 1 \leq i < 8\}$. Observe that ${p_1}^{-1}p_2$ takes $p_1$ to $p_2$, i.e., $p_1({p_1}^{-1}p_2)=(p_1{p_1}^{-1})p_2=p_2$. Equivalently, it is viewed as taking $p_2^{-1}p_1$ to the identity permutation $I$, i.e., $p_2^{-1}p_1({p_1}^{-1}p_2)=I$. Thus, a minimum-length permutation factorization of ${p_1}^{-1}p_2$ using the generating set $S_1^\prime$ corresponds to a shortest path from vertex $p_2^{-1}p_1$ to vertex $I$ in the bubble sort (Cayley) graph $\text{BS}_8$. Therefore, it is reduced to find a shortest path from
$p_2^{-1}p_1=3\,6\,2\,4\,1\,5\,8\,7$ to permutation $I$ in the bubble sort graph $\text{BS}_8$: \\
$3\,6\,2\,4\,1\,5\,8\,7 \xrightarrow{<2, 3>} 3\,2\,6\,4\,1\,5\,8\,7 \xrightarrow{<3, 4>} 3\,2\,4\,6\,1\,5\,8\,7 \xrightarrow{<4, 5>} 3\,2\,4\,1\,6\,5\,8\,7 \xrightarrow{<5, 6>}\\  3\,2\,4\,1\,5\,6\,8\,7 \xrightarrow{<7, 8>} 3\,2\,4\,1\,5\,6\,7\,8 \xrightarrow{<1, 2>} 2\,3\,4\,1\,5\,6\,7\,8 \xrightarrow{<3, 4>} 2\,3\,1\,4\,5\,6\,7\,8\xrightarrow{<2, 3>}\\ 2\,1\,3\,4\,5\,6\,7\,8 \xrightarrow{<1, 2>} 1\,2\,3\,4\,5\,6\,7\,8$, where the label of each arrow denotes an edge in $\text{BS}_8$.  Therefore,  ${p_1}^{-1}p_2={(p_2^{-1}p_1)}^{-1}=(2\,3)(3\,4)(4\,5)(5\,6)(7\,8)(1\,2)(3\,4)(2\,3)(1\,2)$. It follows that the resulting minimum-length sequence of adjacent task swappings needed from $\Gamma_{p_1}^{L(8)}$ to reach $\Gamma_{p_2}^{L(8)}$ is $((2\,3)$, $(3\,4)$, $(4\,5)$, $(5\,6)$, $(7\,8)$, $(1\,2)$, $(3\,4)$, $(2\,3)$, $(1\,2))$. It is interpreted as a sequence of task swappings so that a task swapping between agents in the first term (agent 2 and agent 3) is followed by a task swapping between agents in the second term (agent 3 and agent 4), and so on, until arriving at a task swapping between agents in the last term (agent 1 and agent 2) of the sequence. We see that the length of the sequence is 9, implying that at least 9 adjacent task swappings are needed from $\Gamma_{p_1}^{L(8)}$ to reach $\Gamma_{p_2}^{L(8)}$.  Algorithm~\ref{Algorithm:TSGLine} describes the procedure of converting $\Gamma_{\pi_1}^{L(n)}$ into $\Gamma_{\pi_2}^{L(n)}$ by using a minimum-length sequence of adjacent task swappings. It is known from group theory that the minimum length of permutation $p \in \mathfrak{S}_n$ using the generating set $S_1$ is the \emph{inversion number}~\cite{Garsia2002, Knuth1973} of $p$, where the inversion number of $p$ is defined as $|\{(i, j): i < j, p(i) > p(j)\}|$. The maximum inversion number of permutations of $n$ elements is $n(n-1)/2$, which corresponds to permutation $n\,n-1\,\cdots\,2\,1$~\cite{Garsia2002}. Now, the minimum number of adjacent task swappings needed from $\Gamma_{\pi_1}^{L(n)}$ for $\pi_1 \in \mathfrak{S}_n$ to reach $\Gamma_{\pi_2}^{L(n)}$ for $\pi_2 \in \mathfrak{S}_n$ is the inversion number of ${\pi_1}^{-1}\pi_2$ (or ${\pi_2}^{-1}\pi_1)$. It follows that the least upper bound of the minimum number of adjacent task swappings needed from $\Gamma_{\pi_1}^{L(n)}$ to reach $\Gamma_{\pi_2}^{L(n)}$ is $n(n-1)/2$, which is the maximum inversion number of permutations of $n$ elements. We also see that it coincides the diameter of bubble sort Cayley graph $\text{BS}_n$.

\begin{algorithm}[h!]
\SetAlgoLined
\KwIn{A source and a target task assignment in a task swapping graph of line topology $\Gamma_{\pi_1}^{L(n)}$ and $\Gamma_{\pi_2}^{L(n)}$, respectively.}
\KwOut{A minimum-length sequence of adjacent task swappings needed from $\Gamma_{\pi_1}^{L(n)}$ to reach $\Gamma_{\pi_2}^{L(n)}$.}
\Begin
{
Find a minimum-length permutation factorization of $\pi_1^{-1}\pi_2$ using a shortest path from vertex $\pi_2^{-1}\pi_1$ to vertex $I$ in the bubble-sort Cayley graph $\text{BS}_n$\;
Obtain a minimum-length sequence of adjacent task swappings needed from $\Gamma_{\pi_1}^{L(n)}$ to reach $\Gamma_{\pi_2}^{L(n)}$ by using the above permutation factorization of $\pi_1^{-1}\pi_2$\;
}
\caption{A task reassignment by using adjacent task swappings in a task swapping graph of line topology.}
\label{Algorithm:TSGLine}
\end{algorithm}

In Algorithm~\ref{Algorithm:TSGLine}\footnote{An alternative method to find a minimum-length permutation factorization of permutation $\pi \in \mathfrak{S}_n$ using the generating set $S_1=\{(i\,i+1) : 1 \leq i < n\}$ is to apply \emph{numbers game}~\cite{Bjorner2005, Eriksson1994} of finite Coxeter group of type $A_{n-1}$~\cite{Tenner2006, Garsia2002}. An interested reader may refer to~\cite{Bjorner2005, Eriksson1994, Tenner2006, Garsia2002} for further details.}, we do not need to generate every bubble-sort Cayley graph $\text{BS}_n$ to find a shortest path from vertex $\pi_2^{-1}\pi_1$ to vertex $I$. If a right multiplication of $\pi_2^{-1}\pi_1$ by an adjacent transposition reduces an inversion number by 1, the vertex of the resulting permutation comes closer to vertex $I$ in terms of a distance in $\text{BS}_n$ in which permutation $I$ has the 0 inversion number~\cite{Garsia2002,Akers1989}. By swapping adjacent out-of-order elements in the permutation using the bubble-sort algorithm, it reduces an inversion number by 1. Therefore, we may apply a bubble-sort algorithm of $O(n^2)$ complexity~\cite{Knuth1973} to $\pi_2^{-1}\pi_1$ in order to keep track of a shortest path from vertex $\pi_2^{-1}\pi_1$ to vertex $I$ as shown by the example in this section. Note that a minimum-length sequence of adjacent task swappings needed from $\Gamma_{\pi_1}^{L(n)}$ to reach $\Gamma_{\pi_2}^{L(n)}$ is not necessarily unique. However, the length of a minimum-length sequence is unique, which is the inversion number of permutation $\pi_2^{-1}\pi_1$. It is also the distance from vertex $\pi_2^{-1}\pi_1$ to vertex $I$ in $\text{BS}_n$.

\begin{proposition}
An upper bound for the minimum number of adjacent task swappings needed from $\Gamma_{\pi_1}^{L(n)}$ for $\pi_1 \in \mathfrak{S}_n$ to reach $\Gamma_{\pi_2}^{L(n)}$ for $\pi_2 \in \mathfrak{S}_n$ is $n(n-1)/2$.
\end{proposition}
\begin{proof}
It follows directly from the diameter of Cayley graph $\text{BS}_n$~\cite{Heydemann1997, Lakshmivarahan1993} and from Algorithm~\ref{Algorithm:TSGLine} in which finding a minimum-length permutation factorization of $\pi_1^{-1}\pi_2$ for $\pi_1, \pi_2 \in \mathfrak{S}_n$ using the generating set $S_1=\{(i\,i+1):1 \leq i < n\}$ is converted to the context of finding a minimum-length sequence of adjacent task swappings needed from $\Gamma_{\pi_1}^{L(n)}$ to reach $\Gamma_{\pi_2}^{L(n)}$. $\qquad$
\end{proof}

We next discuss a task swapping graph of star topology. A star topology consists of a supervisor agent and worker agents,  where a supervisor agent communicates directly to worker agents and each worker agent communicates indirectly to other worker agent(s)~\cite{Zhu2006, Zhang2006, El-Rewini1998}. We assume that a task swapping graph of star topology having $n\,(n \geq 3)$ agents is labeled in such a way that a supervisor agent is labeled 1 and worker agents are labeled  in ascending order (clockwise) starting from 2 to $n$. A task swapping graph of star topology having $n$ agents with their task assignment represented by permutation $p \in \mathfrak{S}_n$ is denoted as $\Gamma_p^{S(n)}$. For example, a task swapping graph of Figure~\ref{fig:StarTopology}(a) is denoted as $\Gamma_{p_1}^{S(9)}$ and a task swapping graph of Figure~\ref{fig:StarTopology}(b) is denoted as $\Gamma_{p_2}^{S(9)}$, respectively,  where $p_1=5\,4\,2\,1\,6\,9\,7\,8\,3 \in \mathfrak{S}_9$ and $p_2=7\,8\,9\,2\,4\,5\,1\,6\,3 \in \mathfrak{S}_9$. Let $p \in \mathfrak{S}_n$ be a permutation representing a task assignment in $\Gamma_{p}^{S(n)}$. We see that a right multiplication of permutation $p$ by a \emph{star transposition}~\cite{Pak1999, Irving2009} $(1\,i)$ for $2 \leq i \leq n$ exchanges the values in position $1$ (a supervisor agent's position) of $p$ and position $i$ (a worker agent's position) of $p$. Therefore, to find a minimum-length sequence of adjacent task swappings needed from $\Gamma_{\pi_1}^{S(n)}$ for $\pi_1 \in \mathfrak{S}_n$ to reach $\Gamma_{\pi_2}^{S(n)}$ for $\pi_2 \in \mathfrak{S}_n$ is equivalent to finding a minimum-length permutation factorization of $\pi_1^{-1}\pi_2$ using the generating set $S_3=\{(1\,i) : 2 \leq i \leq n\}$. Observe that every non-identity permutation is denoted as a product of disjoint cycles, one of which includes element 1. Therefore, a non-identity permutation $\pi \in \mathfrak{S}_n$ is expressed as $\pi=(1\,q_2 \cdots q_{s})(p_1^1 \cdots p_{l_1}^1) \cdots (p_1^m \cdots p_{l_m}^m) \in \mathfrak{S}_n$ if $s \geq 2$~\cite{Irving2009}. If a cycle including element 1 is a cycle of length 1, then permutation $\pi \in \mathfrak{S}_n$ is expressed as $\pi=(p_1^1 \cdots p_{l_1}^1) \cdots (p_1^m \cdots p_{l_m}^m) \in \mathfrak{S}_n$. It is easily verified that $(1\,q_2 \cdots q_{s})$ in $\pi$ for $s \geq 2$ is factorized into $(1\,q_{s})(1\,q_{s-1}) \cdots (1\,q_2)$ using the generating set $S_3$, whose length is $s -1$. Similarly, if $l_k \geq 2$, then $(p_1^k \cdots p_{l_k}^k)$ in $\pi$ is factorized into $(1\,p_1^k)$$(1\,p_{l_k}^k)(1\,p_{l_k-1}^k) \cdots (1\,p_1^k)$ of length $l_k+ 1$ using the generating set $S_3$~\cite{Irving2009}. It turns out these factorizations are minimal in the sense that no other way of factorizations can have less length when using the generating set $S_3$~\cite{Irving2009}. Therefore, the minimum length of the above $\pi \in \mathfrak{S}_n$ using the generating set $S_3$ is $n+m-k-1$, where $k$ is the number of cycle(s) of length 1. For example, if $\pi=(1\,2\,3\,4)(5\,6\,7) \in \mathfrak{S}_8$, then $\pi$ is factorized into $\pi=(1\,4)(1\,3)(1\,2)(1\,5)(1\,7)(1\,6)(1\,5)$ using the generating set $S_3^\prime=\{(1\,i) : 2 \leq i \leq 8\}$, whose length is $8+1-1-1=7$.

\begin{algorithm}[h!]
\SetAlgoLined
\KwIn{A source and a target task assignment in a task swapping graph of star topology $\Gamma_{\pi_1}^{S(n)}$ and $\Gamma_{\pi_2}^{S(n)}$, respectively.}
\KwOut{A minimum-length sequence of adjacent task swappings needed from $\Gamma_{\pi_1}^{S(n)}$ to reach $\Gamma_{\pi_2}^{S(n)}$.}
\Begin
{
Compute $\pi_1^{-1}\pi_2$ and set it as $\pi$\;
Denote $\pi$ as a product of disjoint cycles such that $(1\,q_2 \cdots q_{s})(p_1^1 \cdots p_{l_1}^1) \cdots (p_1^m \cdots p_{l_m}^m) \in \mathfrak{S}_n$ if $s \geq 2$.  Denote $\pi$ as $(p_1^1 \cdots p_{l_1}^1) \cdots (p_1^m \cdots  p_{l_m}^m) \in \mathfrak{S}_n$ otherwise\;
\If{$s \geq 3$}
{
$(1\,q_2 \cdots q_{s})$ in $\pi$ is factorized into $(1\,q_{s})(1\,q_{s-1}) \cdots (1\,q_2)$\;
}

\For{$k\leftarrow 1$ \KwTo $m$}
{
\If{$l_k \geq 2$}
{
$(p_1^k \cdots p_{l_k}^k)$ in $\pi$ is factorized into $(1\,p_1^k)$$(1\,p_{l_k}^k)(1\,p_{l_k-1}^k) \cdots (1\,p_1^k)$\;
}
}
Obtain a minimum-length sequence of adjacent task swappings needed from $\Gamma_{\pi_1}^{S(n)}$ to reach $\Gamma_{\pi_2}^{S(n)}$ by using the above permutation factorization of $\pi_1^{-1}\pi_2$\;
}
\caption{A task reassignment by using adjacent task swappings in a task swapping graph of star topology.}
\label{Algorithm:TSGStar}
\end{algorithm}

Algorithm~\ref{Algorithm:TSGStar} describes the procedure of converting $\Gamma_{\pi_1}^{S(n)}$ into $\Gamma_{\pi_2}^{S(n)}$ by using a minimum-length sequence of adjacent task swappings. For example, we find a minimum-length sequence of adjacent task swappings needed from $\Gamma_{p_1}^{S(9)}$ to reach $\Gamma_{p_2}^{S(9)}$ in Figure~\ref{fig:StarTopology}, where $p_1=5\,4\,2\,1\,6\,9\,7\,8\,3 \in \mathfrak{S}_9$ and $p_2=7\,8\,9\,2\,4\,5\,1\,6\,3 \in \mathfrak{S}_9$. A simple computation shows that $p_1^{-1}p_2=7\,8\,6\,3\,2\,1\,4\,5\,9=(1\,7\,4\,3\,6)(2\,8\,5) \in \mathfrak{S}_9$. By applying Algorithm~\ref{Algorithm:TSGStar}, we factorize $p_1^{-1}p_2$ into a product of star transpositions, i.e., $p_1^{-1}p_2 = (1\,6)(1\,3)$$(1\,4)$$(1\,7)$$(1\,2)$$(1\,5)$$(1\,8)$$(1\,2)$. Now, a minimum-length sequence of adjacent task swappings needed from $\Gamma_{p_1}^{S(9)}$ to reach $\Gamma_{p_2}^{S(9)}$ is  $((1\,6)$, $(1\,3)$, $(1\,4)$, $(1\,7)$, $(1\,2)$, $(1\,5)$, $(1\,8)$, $(1\,2))$ of length 8.

Observe that a sequence of adjacent task swappings needed from $\Gamma_{\pi_1}^{S(n)}$ to reach $\Gamma_{\pi_2}^{S(n)}$ in Algorithm~\ref{Algorithm:TSGStar} corresponds to a path from vertex $\pi_2^{-1}\pi_1$ to vertex $I$ in a star graph $\text{ST}_n$. Therefore, an upper bound of the minimum number of adjacent task swappings needed from $\Gamma_{\pi_1}^{S(n)}$ to reach $\Gamma_{\pi_2}^{S(n)}$ in Algorithm~\ref{Algorithm:TSGStar} is the diameter of a star graph $\text{ST}_n$, which is $\lfloor3(n-1)/2\rfloor$~\cite{Akers1989, Lakshmivarahan1993}.

\begin{proposition}
An upper bound for the minimum number of adjacent task swappings needed from $\Gamma_{\pi_1}^{S(n)}$ for $\pi_1 \in \mathfrak{S}_n$ to reach $\Gamma_{\pi_2}^{S(n)}$ for $\pi_2 \in \mathfrak{S}_n$ is $\lfloor3(n-1)/2\rfloor$.
\end{proposition}
\begin{proof}
It follows immediately from the diameter of Cayley graph $\text{ST}_n$~\cite{Lakshmivarahan1993, Akers1989} and from Algorithm~\ref{Algorithm:TSGStar} in which finding a minimum-length permutation factorization of $\pi_1^{-1}\pi_2$  for $\pi_1, \pi_2 \in \mathfrak{S}_n$ using the generating set $S_3=\{(1\,i):2 \leq i \leq n\}$ is converted to the context of finding a minimum-length sequence of adjacent task swappings needed from $\Gamma_{\pi_1}^{S(n)}$ to reach $\Gamma_{\pi_2}^{S(n)}$. $\qquad$
\end{proof}

\begin{figure}[h!]
\centering
\includegraphics[width=0.99\textwidth]{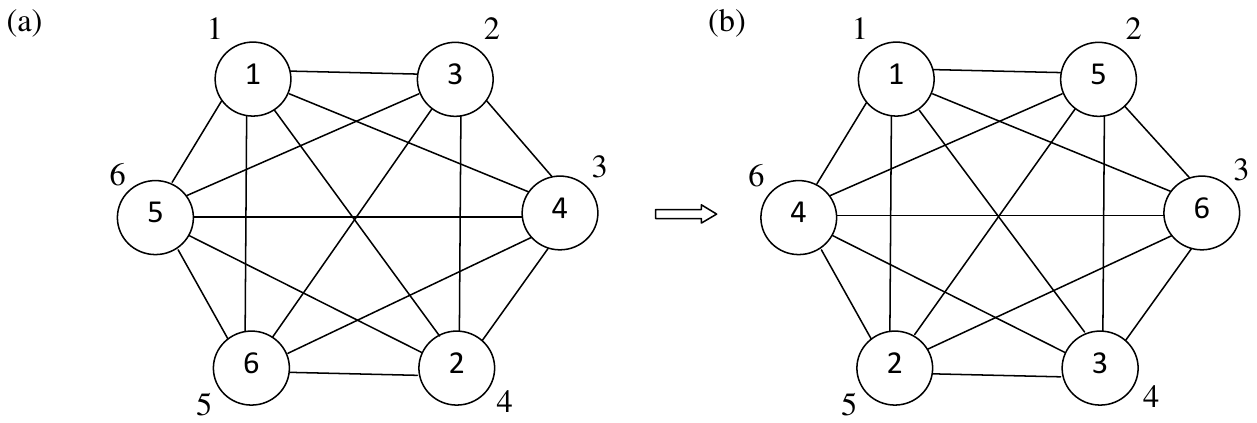}
\caption{Task swapping graphs of complete topology.}
\label{fig:CompleteTopology}
\end{figure}

A task swapping graph of complete topology is a fully-connected task swapping graph in which each agent has direct links with all other agents in the topology. Although the complete topology provides redundancy in terms of communication links between pairs of agents, the cost is often too high when setting up communication links between agents in the topology (i.e., $n(n-1)/2$ total communication links are required for $n$ agents in the complete topology)~\cite{Zhu2006, Zhang2006}. A task swapping graph of complete topology having $n$ agents with their task assignment represented by permutation $p \in \mathfrak{S}_n$ is denoted as $\Gamma_p^{C(n)}$. For example, a task swapping graph of Figure~\ref{fig:CompleteTopology}(a) is denoted as $\Gamma_{p_1}^{C(6)}$ and a task swapping graph of Figure~\ref{fig:CompleteTopology}(b) is denoted as $\Gamma_{p_2}^{C(6)}$, respectively,  where $p_1=1\,3\,4\,2\,6\,5 \in \mathfrak{S}_6$ and $p_2=1\,5\,6\,3\,2\,4 \in \mathfrak{S}_6$. Since each agent has direct links with all other agents in the complete topology, an adjacent task swapping may occur between any pair of agents in the topology. Now, to find a minimum-length sequence of adjacent task swappings needed from $\Gamma_{\pi_1}^{C(n)}$ to reach $\Gamma_{\pi_2}^{C(n)}$ is reduced to find a minimum-length permutation factorization of $\pi_1^{-1}\pi_2$ using the generating set $S_4=\{(i\,j):1 \leq i <j \leq n\}$. Verify that each cycle $(q_1\, q_2 \cdots q_l)$ of length $ l > 2$ can be factorized into a product of $l-1$ transpositions $(q_1\,q_l)(q_1\,q_{l-1}) \cdots (q_1\,q_2)$. It is known from group theory that a cycle of length $l > 2$ cannot be written as a product of fewer than $l-1$ transpositions in the generating set $S_4$~\cite{Neuenschwander2001, Mackiw1995}. Thus, a minimum-length permutation factorization of $\pi_1^{-1}\pi_2$ is obtained by first denoting it as a product of disjoint cycles, then factorizing all the cycle(s) of length greater than 2 using the generating set $S_4$ as described. 

\begin{algorithm}[h!]
\SetAlgoLined
\KwIn{A source and a target task assignment in a task swapping graph of complete topology $\Gamma_{\pi_1}^{C(n)}$ and $\Gamma_{\pi_2}^{C(n)}$, respectively.}
\KwOut{A minimum-length sequence of adjacent task swappings needed from $\Gamma_{\pi_1}^{C(n)}$ to reach $\Gamma_{\pi_2}^{C(n)}$.}
\Begin
{
Compute $\pi_1^{-1}\pi_2$ and set it as $\pi$\;
Denote $\pi$ as a product of disjoint cycles such that $(q_1^1\,\cdots q_{l_1}^1)(q_1^2 \cdots q_{l_2}^2) \cdots (q_1^m \cdots q_{l_m}^m) \in \mathfrak{S}_n$\;
\For{$k\leftarrow 1$ \KwTo $m$}
{
\If{$l_k \geq 3$}
{
 $(q_1^k\,q_2^k \cdots q_l^k)$ in $\pi$ is factorized into $(q_1^k\,q_l^k)\cdots (q_1^k\,q_2^k)$\;
}
}
Obtain a minimum-length sequence of adjacent task swappings needed from $\Gamma_{\pi_1}^{C(n)}$ to reach $\Gamma_{\pi_2}^{C(n)}$ by using the above permutation factorization of $\pi_1^{-1}\pi_2$\;
}
\caption{A task reassignment by using adjacent task swappings in a task swapping graph of complete topology.}
\label{Algorithm:TSGComplete}
\end{algorithm}

Algorithm~\ref{Algorithm:TSGComplete} describes the procedure of converting $\Gamma_{\pi_1}^{C(n)}$ into $\Gamma_{\pi_2}^{C(n)}$ by using a minimum-length sequence of adjacent task swappings. We see that a minimum length of permutation $\pi \in \mathfrak{S}_n$ using the generating set $S_4$ is $n-r$, where $\pi$ consists of $r$ disjoint cycles.

Now, we find a minimum-length sequence of adjacent task swappings needed from $\Gamma_{p_1}^{C(6)}$ to reach $\Gamma_{p_2}^{C(6)}$ in Figure~\ref{fig:CompleteTopology}, where $p_1=1\,3\,4\,2\,6\,5 \in \mathfrak{S}_6$ and $p_2=1\,5\,6\,3\,2\,4 \in \mathfrak{S}_6$. A direct computation shows that $p_1^{-1}p_2=1\,6\,5\,2\,4\,3 = (2\,6\,3\,5\,4) \in \mathfrak{S}_6$. Then, we factorize $p_1^{-1}p_2$ as a product of transpositions, i.e., $p_1^{-1}p_2 = (2\,4)$$(2\,5)$$(2\,3)$$(2\,6)$ by applying Algorithm~\ref{Algorithm:TSGComplete}. Therefore, a minimum-length sequence of adjacent task swappings needed from $\Gamma_{p_1}^{C(6)}$ to reach $\Gamma_{p_2}^{C(6)}$ is $((2\,4)$, $(2\,5)$, $(2\,3)$, $(2\,6))$ of length 4. 

In Section~\ref{sec:TranspositionGraphs} we discussed that a Cayley graph of $\mathfrak{S}_n$ generated by $S_4=\{(i\,j):1 \leq i <j \leq n\}$ is the complete transposition graph $\text{CT}_n$.  It follows that a sequence of adjacent task swappings needed from $\Gamma_{\pi_1}^{C(n)}$ to reach $\Gamma_{\pi_2}^{C(n)}$ in Algorithm~\ref{Algorithm:TSGComplete} corresponds to a path from vertex $\pi_2^{-1}\pi_1$ to vertex $I$ in the complete transposition graph $\text{CT}_n$. To find a shortest path from vertex
$\pi_2^{-1}\pi_1$ to vertex $I$ in $\text{CT}_n$~\cite{Lakshmivarahan1993}, one may apply a greedy algorithm to transpose and locate each element in the permutation to its homed position iteratively from left to right until arriving at $I$. For example, a shortest path from vertex $p_2^{-1}p_1={(p_1^{-1}p_2)}^{-1}=1\,4\,6\,5\,3\,2 \in \mathfrak{S}_6$ to vertex $I$ in $\text{CT}_6$ is as follows: $1\,4\,6\,5\,3\,2 \xrightarrow{<2, 4>} 1\,5\,6\,4\,3\,2 \xrightarrow{<2, 5>} 1\,3\,6\,4\,5\,2$$ \xrightarrow{<2, 3>} 1\,6\,3\,4\,5\,2 \xrightarrow{<2, 6>} 1\,2\,3\,4\,5\,6$, where the label of each arrow denotes an edge in $\text{CT}_6$.  Therefore,  ${p_1}^{-1}p_2={(p_2^{-1}p_1)}^{-1}=(2\,4)$$(2\,5)$$(2\,3)$$(2\,6)$, which coincides the above permutation factorization of ${p_1}^{-1}p_2$ using the generating set $S_4$. 

It follows that an upper bound for the minimum number of adjacent task swappings needed from $\Gamma_{\pi_1}^{C(n)}$ to reach $\Gamma_{\pi_2}^{C(n)}$ in Algorithm~\ref{Algorithm:TSGComplete} is the diameter of a complete transposition graph $\text{CT}_n$, which is $n-1$~\cite{Lakshmivarahan1993}.

\begin{proposition}
An upper bound for the minimum number of adjacent task swappings needed from $\Gamma_{\pi_1}^{C(n)}$ for $\pi_1 \in \mathfrak{S}_n$ to reach $\Gamma_{\pi_2}^{C(n)}$ for $\pi_2 \in \mathfrak{S}_n$ is $n-1$.
\end{proposition}
\begin{proof}
It follows directly from the diameter of Cayley graph $\text{CT}_n$~\cite{Lakshmivarahan1993, Heydemann1997} and from Algorithm~\ref{Algorithm:TSGComplete} in which finding a minimum-length permutation factorization of $\pi_1^{-1}\pi_2$ for $\pi_1, \pi_2 \in \mathfrak{S}_n$ using the generating set $S_4=\{(i\,j):1 \leq i <j \leq n\}$ is converted to the context of finding a minimum-length sequence of adjacent task swappings needed from $\Gamma_{\pi_1}^{C(n)}$ to reach $\Gamma_{\pi_2}^{C(n)}$. $\qquad$
\end{proof}

\begin{figure}[h!]
\centering
\includegraphics[width=0.99\textwidth]{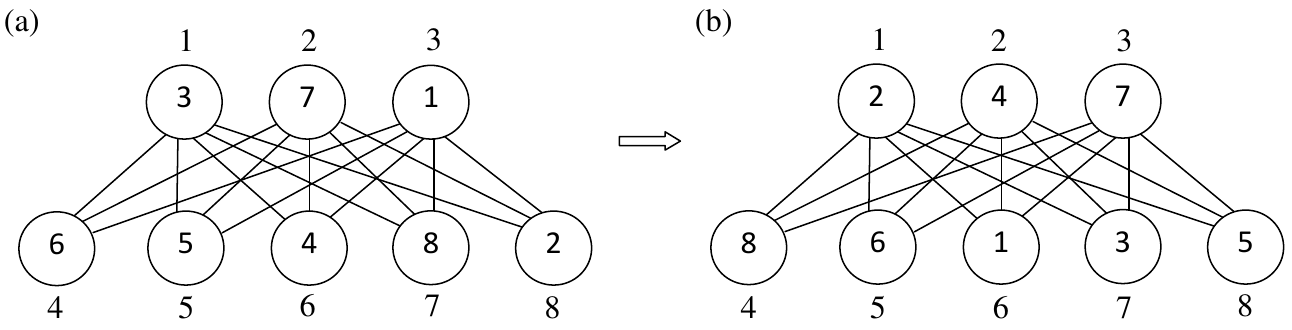}
\caption{Task swapping graphs of complete bipartite topology.}
\label{fig:BipartiteTopology}
\end{figure}

Recall that a graph is \emph{bipartite} if its vertex set admits a partition into two classes in such a way that every edge has its ends in two different classes~\cite{Biggs1974, Diestel2005}. A \emph{complete bipartite graph}~\cite{Diestel2005} is a bipartite graph in which every two vertices from two different classes are adjacent. A task swapping graph of \emph{complete bipartite topology}~\cite{Zhu2006, Zhu1996} is a complete bipartite task swapping graph, where agents are grouped in two layers or classes (the upper and lower) in such a way that agents between different layers are fully-connected, while agents in the same layer are not directly connected. The agents at the upper layer have a centralized control over the agents at the lower layer in a distributed manner in this topology. We assume that each agent in a task swapping graph of complete bipartite topology is labeled in ascending order from the upper left to bottom right as shown in Figure~\ref{fig:BipartiteTopology}. A task swapping graph of complete bipartite topology having $n$ agents with their task assignment represented by permutation $p \in \mathfrak{S}_n$ is denoted as $\Gamma_p^{B(n, k)}$, where $k$ is the number of agents at the upper layer in $\Gamma_p^{B(n, k)}$. We call $k$ as \emph{bipartite index}. As with other task swapping graphs, we assume $n \geq 3$ for $\Gamma_p^{B(n, k)}$. We also assume $1 \leq k <n$ for $\Gamma_p^{B(n, k)}$. Note that $\Gamma_p^{B(n, k)}$ is simply $\Gamma_p^{S(n)}$ when bipartite index $k$ is 1. For example, the task swapping graph of Figure~\ref{fig:BipartiteTopology}(a) is denoted as $\Gamma_{p_1}^{B(8, 3)}$ and the task swapping graph of Figure~\ref{fig:BipartiteTopology}(b) is denoted as $\Gamma_{p_2}^{B(8, 3)}$, respectively,  where $p_1=3\,7\,1\,6\,5\,4\,8\,2 \in \mathfrak{S}_8$ and $p_2=2\,4\,7\,8\,6\,1\,3\,5 \in \mathfrak{S}_8$. We see that a right multiplication of permutation $p_1 \in \mathfrak{S}_8$ by transposition $(1\,4)$ represents an adjacent task swapping between agent 1 and agent 4 in $\Gamma_{p_1}^{B(8, 3)}$, while a right multiplication of permutation $p_1 \in \mathfrak{S}_8$ by transposition $(1\,2)$ does not represent an adjacent task swapping in $\Gamma_{p_1}^{B(8, 3)}$. Therefore, the generating set $S_5^\prime=\{(i\,j):1 \leq i \leq 3 <j \leq 8\}$ is required for finding a minimum-length sequence of adjacent task swappings needed from $\Gamma_{p_1}^{B(8, 3)}$ to reach $\Gamma_{p_2}^{B(8, 3)}$. 

We next find a minimum-length permutation factorization of $\pi_1^{-1}\pi_2$ using the generating set $S_5=\{(i\,j):1 \leq i \leq k <j \leq n\}$ by which we obtain a minimum-length sequence of adjacent task swappings needed from $\Gamma_{\pi_1}^{B(n, k)}$ for $\pi_1 \in \mathfrak{S}_n$ to reach $\Gamma_{\pi_2}^{B(n, k)}$ for $\pi_2 \in \mathfrak{S}_n$. We consider several types of cycles to factorize $\pi_1^{-1}\pi_2$ using the generating set $S_5=\{(i\,j):1 \leq i \leq k <j \leq n\}$. Set $\pi=\pi_1^{-1}\pi_2$ and represent it as a product of (commutative) disjoint cycles $C_s$ for $1 \leq s \leq t$ such that $\pi=C_1 \cdots C_i C_{i+1} \cdots C_e C_{e+1} \cdots C_t$, where each element in each cycle of $C_1 \cdots C_i$ is less than or equal to (bipartite index) $k$, each element in each cycle of $C_{i+1} \cdots C_e$ is greater than $k$, and each cycle of $C_{e+1} \cdots C_t$ has both element(s) less than or equal to $k$ and element(s) greater than $k$. We first consider the first type of a cycle, referred to as an \emph{internal cycle}~\cite{Zhu1996}, which is a cycle in $C_1 \cdots C_i$ of $\pi$. Let $C_x$ be an internal cycle such that $C_x = (c_1\,c_2\,\cdots\,c_v)$, where $1\leq c_u \leq k$ for $1 \leq u \leq v$. Then, $C_x$ is factorized into $(c_1\,t)(c_v\,t)\cdots(c_2\,t)(c_1\,t)$ using the generating set $S_5=\{(i\,j) : 1 \leq i \leq k < j \leq n\}$, where $t$ is an arbitrary number satisfying $k < t \leq n$. We next consider the second type of a cycle, referred to as an \emph{external cycle}~\cite{Zhu1996}, which is a cycle in $C_{i+1} \cdots C_e$ of $\pi$. Let $C_y$ be an external cycle, i.e., $C_y = (c_1\,\cdots\,c_{q-1}\,c_q)$, where $k < c_p \leq n$ for $1 \leq p \leq q$. Then, $C_y = (c_1\,\cdots\,c_{q-1}\,c_q)$ is factorized into $(u\,c_q)(u\,c_{q-1})\cdots(u\,c_1)(u\,c_q)$ using the generating set $S_5$, where $u$ is an arbitrary number satisfying $1 \leq u \leq k$. Now, we consider the final type of a cycle, referred to as a \emph{mixed cycle}~\cite{Zhu1996}, which is a cycle in $C_{e+1} \cdots C_t$ of $\pi$. Each mixed cycle $C_m$ for $e+1 \leq m \leq t$ is written as $(E_{m_1}E_{m_2}\cdots E_{m_s})$, where each $E_{m_i}$ for $1 \leq i \leq s$ can be denoted as a concatenation of two blocks of numbers~\cite{Zhu1996}. Each number of the first block of $E_{m_i}$ for $1 \leq i \leq s$ is less than or equal to $k$, while each number of the second block of $E_{m_i}$ for $1 \leq i \leq s$ is greater than $k$. We call $E_{m_i}$ as a \emph{simple cycle}. For example, a mixed cycle $C_m^\prime = (1\,3\,4\,5\,7\,2\,6\,8) \in \mathfrak{S}_8$ for $k=3$ is written as $(E_{m_1}^\prime E_{m_2}^\prime)$, where $E_{m_1}^\prime= 1\,3\,4\,5\,7$ and $E_{m_2}^\prime= 2\,6\,8$. It follows that the first block of  $E_{m_1}^\prime$ is $1\,3$, while the second block of $E_{m_1}^\prime$ is $4\,5\,7$. In general, let $E_{m_i}=i_1\,i_2\cdots i_a\,j_1\,j_2\cdots j_b$, where $1 \leq i_u \leq k$ for $1 \leq u \leq a$ and $k < j_v \leq n$ for $1 \leq v \leq b$. Observe that the cycle $(E_{m_i})$ is factorized into $(i_1\,j_b)\cdots(i_1\,j_2)(i_a\,j_1)\cdots(i_2\,j_1)(i_1\,j_1)$ using the generating set $S_5$. For example, $(1\,3\,4\,5\,7) \in \mathfrak{S}_8$ is factorized into $(1\,7)(1\,5)(3\,4)(1\,4)$ using the generating set $S_5^\prime=\{(i\,j):1 \leq i \leq 3 <j \leq 8\}$. Further, observe that a mixed cycle $C_m^\prime=(E_{m_1}^\prime E_{m_2}^\prime)$ is written as $(1\,2)(E_{m_1}^\prime)(E_{m_2}^\prime)$, where 1 in $(1\,2)$ is the first element in $E_{m_1}^\prime$ and 2 in $(1\,2)$ is the first element in $E_{m_2}^\prime$. Thus, $C_m^\prime=(E_{m_1}^\prime E_{m_2}^\prime) = (1\,2)(E_{m_1}^\prime)(E_{m_2}^\prime) = (1\,2)(1\,7)(1\,5)(3\,4)(1\,4)(2\,8)(2\,6)$. However, $(1\,2)$ in $C_m^\prime$ is not a transposition in $S_5^\prime$. Therefore, we use a transposition $(1\,7)$ next to $(1\,2)$ and convert $(1\,2)(1\,7)$ into $(1\,7)(2\,7)$ in which transposition $(1\,7)$ and transposition $(2\,7)$ are transpositions in $S_5^\prime$. In general $C_m = (E_{m_1}E_{m_2}\cdots E_{m_s})$ is recursively factorized into $C_m = (v_1\,v_2)(E_{m_1})(E_{m_2}\cdots E_{m_s})$, where $v_1$ is the first element of $E_{m_1}$ and $v_2$ is the first element of $E_{m_2}$~\cite{Zhu1996}. Let $(w_1\,w_2)$ be the first transposition of a factorization of $E_{m_1}$ using the generating set $S_5$. Then, we have $v_1 = w_1$. Now, we see that $E_{m_1}$ is factorized using transpositions in the generating set $S_5$ such that $(v_1\,v_2)(w_1\,w_2)$ with $v_1=w_1$ is rearranged into $(v_1\,w_2)(v_2\,w_2)$. Therefore, $C_m$ is recursively factorized using transpositions in the generating set $S_5$. Furthermore, it turns out that the above way of factorizing an arbitrary permutation $p \in \mathfrak{S}_n$ using the generating set $S_5$ for a given $1 \leq k < n$ yields the length of $p$ not greater than $n-1+ \max{(\lfloor k/2  \rfloor,  \lfloor (n-k)/2 \rfloor)}$~\cite{Zhu1996}. Algorithm~\ref{Algorithm:TSGBipartite} describes the procedure of converting $\Gamma_{\pi_1}^{B(n, k)}$ into $\Gamma_{\pi_2}^{B(n, k)}$ by using a minimum-length sequence of adjacent task swappings. Now, we obtain a minimum-length sequence of adjacent task swappings needed from a task swapping graph $\Gamma_{p_1}^{B(8, 3)}$ in Figure~\ref{fig:BipartiteTopology}(a) to reach a task swapping graph $\Gamma_{p_2}^{B(8, 3)}$ in Figure~\ref{fig:BipartiteTopology}(b), where $p_1=3\,7\,1\,6\,5\,4\,8\,2 \in \mathfrak{S}_8$ and $p_2=2\,4\,7\,8\,6\,1\,3\,5 \in \mathfrak{S}_8$. We first compute $p_1^{-1}p_2$, which is $8\,6\,2\,7\,4\,3\,1\,5 = (1\,8\,5\,4\,7)(3\,2\,6) \in \mathfrak{S}_8$. Then, we factorize $p_1^{-1}p_2$ using the generating set $S_5^\prime$. Observe that $p_1^{-1}p_2$ is the product of simple cycles $(E_1)(E_2)$, where $E_1 = 1\,8\,5\,4\,7$ and $E_2 = 3\,2\,6$. Then, $(E_1) = (1\,7)(1\,4)(1\,5)(1\,8)$ and $(E_2)=(2\,6)(3\,6)$. Therefore, a minimum-length sequence of adjacent task swapping needed from  $\Gamma_{p_1}^{B(8, 3)}$ in Figure~\ref{fig:BipartiteTopology}(a) to reach $\Gamma_{p_2}^{B(8, 3)}$ in Figure~\ref{fig:BipartiteTopology}(b) is $( (1\,7), (1\,4), (1\,5), (1\,8), (2\,6), (3\,6))$ of length 6.

\begin{algorithm}[h!]
\SetAlgoLined
\KwIn{A source and a target task assignment in a task swapping graph of complete bipartite topology $\Gamma_{\pi_1}^{B(n, k)}$ and $\Gamma_{\pi_2}^{B(n, k)}$, respectively.}
\KwOut{A minimum-length sequence of adjacent task swappings needed from $\Gamma_{\pi_1}^{B(n, k)}$ to reach $\Gamma_{\pi_2}^{B(n, k)}$.}
\Begin
{
Compute $\pi_1^{-1}\pi_2$ and set it as $\pi$\;
Write $\pi$ as a product of disjoint cycles $C_m$ for $1 \leq m \leq t$ such that $\pi=C_1 \cdots C_i C_{i+1} \cdots C_e C_{e+1} \cdots C_t$, where each cycle of $C_1 \cdots C_i$ is an internal cycle, each cycle of $C_{i+1} \cdots C_e$ is an external cycle,  and each cycle of $C_{e+1} \cdots C_t$ is a mixed cycle\; 
\For{$m\leftarrow 1$ \KwTo $t$}
{
\If{$C_m$ is an $r$-cycle for $r \geq 3$}
{
\tcp*[h]{\ see Alogorithm~\ref{Algorithm:BipartiteCycleFactorization}}\break
BipartiteCycleFactorization ($C_m$, $k$)\;
}
}
Obtain a minimum-length sequence of adjacent task swappings needed from $\Gamma_{\pi_1}^{B(n, k)}$ to reach $\Gamma_{\pi_2}^{B(n, k)}$ by using the above permutation factorization of $\pi_1^{-1}\pi_2$\;
}
\caption{A task reassignment by using adjacent task swappings in a task swapping graph of complete bipartite topology.}
\label{Algorithm:TSGBipartite}
\end{algorithm}

\begin{algorithm}[h!]
\SetAlgoLined
\KwIn{An $r$-cycle $C_m$  ($3 \leq r \leq n$) and bipartite index $k$ ($1 \leq k <n$).}
\KwOut{A factorization of $C_m$ into a product of transpositions in $S_5$.}
\Begin
{
\If{$C_m$ is an internal cycle}
{
Let $C_m = (c_1\,c_2\,\cdots\,c_v)$. Then, $C_m$ is factorized into $(c_1\,t)(c_v\,t)\cdots(c_2\,t)(c_1\,t)$, where $t$ is any number in $k < t \leq n$\;
}
\ElseIf{$C_m$ is an external cycle}
{
Let $C_m = (c_1\,\cdots\,c_{q-1}\,c_q)$. Then, $C_m$ is factorized into $(u\,c_q)(u\,c_{q-1})\cdots(u\,c_1)(u\,c_q)$, where $u$ is any number in $1 \leq u \leq k$. 
}
\Else (\tcp*[h]{$C_m$ is a mixed cycle})
{
\If{$C_m$ is a simple cycle}
{
Let $C_m = (i_1\,i_2\cdots i_a\,j_1\,j_2\cdots j_b)$, where $1 \leq i_u \leq k$ for $1 \leq u \leq a$ and $k < j_v \leq n$ for $1 \leq v \leq b$. Then, $C_m$ is factorized into $(i_1\,j_b)\cdots(i_1\,j_2)(i_a\,j_1)\cdots(i_2\,j_1)(i_1\,j_1)$\;
}
\Else
{
Let $C_m= (E_{m_1}E_{m_2}\cdots E_{m_t})$, where each $(E_{m_i})$ for $1 \leq i \leq t$ is a simple cycle. Then, $C_m = (v_1\,v_2)(E_{m_1})(E_{m_2}\cdots E_{m_t})$, where $v_1$ and $v_2$ are the first elements of $E_{m_1}$ and $E_{m_2}$, respectively. Factorize a simple cycle $(E_{m_1})$ as indicated above. Let $(w_1\,w_2)$ be the first transposition of a factorization of $E_{m_1}$. If $v_1 = w_1$, rearrange $(v_1\,v_2)(w_1\,w_2)$ into $(w_1\,w_2)(v_2\,w_2)$. Otherwise, rearrange $(v_1\,v_2)(w_1\,w_2)$ into $(w_1\,w_2)(v_2\,w_2)(v_1\,w_2)(v_2\,w_2)$. Then, $C_m$ is written as a product of transpositions in $S_5$ followed by $(E_{m_2}\cdots E_{m_t})$. We repeat this process recursively to $(E_{m_2}\cdots E_{m_t})$ until we completely factorize $C_m$ using $S_5$.
}
}
}
\caption{A bipartite cycle factorization using the generating set $S_5=\{(i\,j) : 1 \leq i \leq k < j \leq n\}$~\cite{Zhu1996}.}
\label{Algorithm:BipartiteCycleFactorization}
\end{algorithm}

As discussed in Section~\ref{sec:TranspositionGraphs}, a Cayley graph of $\mathfrak{S}_n$ generated by $S_5=\{(i\,j) : 1 \leq i \leq k < j \leq n\}$ is the generalized star graph $\text{GST}_{n, k}$. Therefore, an upper bound of the minimum number of adjacent task swappings needed from $\Gamma_{\pi_1}^{B(n, k)}$ for $\pi_1 \in \mathfrak{S}_n$ to reach $\Gamma_{\pi_2}^{B(n, k)}$ for $\pi_2 \in \mathfrak{S}_n$ in Algorithm~\ref{Algorithm:TSGBipartite} is the diameter of a generalized star graph $\text{GST}_{n,k}$, which is $n-1+ \max{(\lfloor k/2  \rfloor,  \lfloor (n-k)/2 \rfloor)}$~\cite{Zhu1996}.

\begin{proposition}
An upper bound for the minimum number of adjacent task swappings needed from $\Gamma_{\pi_1}^{B(n, k)}$ for $\pi_1 \in \mathfrak{S}_n$ to reach $\Gamma_{\pi_2}^{B(n, k)}$ for $\pi_2 \in \mathfrak{S}_n$ is $n-1+ \max{(\lfloor k/2  \rfloor,  \lfloor (n-k)/2 \rfloor)}$.
\end{proposition}
\begin{proof}
It follows directly from the diameter of Cayley graph $\text{GST}_{n,k}$~\cite{Heydemann1997, Zhu1996}, a bipartite cycle factorization algorithm given in ~\cite{Zhu1996}, and Algorithm~\ref{Algorithm:TSGBipartite} and \ref{Algorithm:BipartiteCycleFactorization} in which finding a minimum-length permutation factorization of $\pi_1^{-1}\pi_2$ for $\pi_1, \pi_2 \in \mathfrak{S}_n$ using the generating set $S_5=\{(i\,j):1 \leq i \leq k <j \leq n\}$ is converted to the context of finding a minimum-length sequence of adjacent task swappings needed from $\Gamma_{\pi_1}^{B(n, k)}$ to reach $\Gamma_{\pi_2}^{B(n, k)}$. $\qquad$
\end{proof}

\begin{figure}[h!]
\centering
\includegraphics[width=0.99\textwidth]{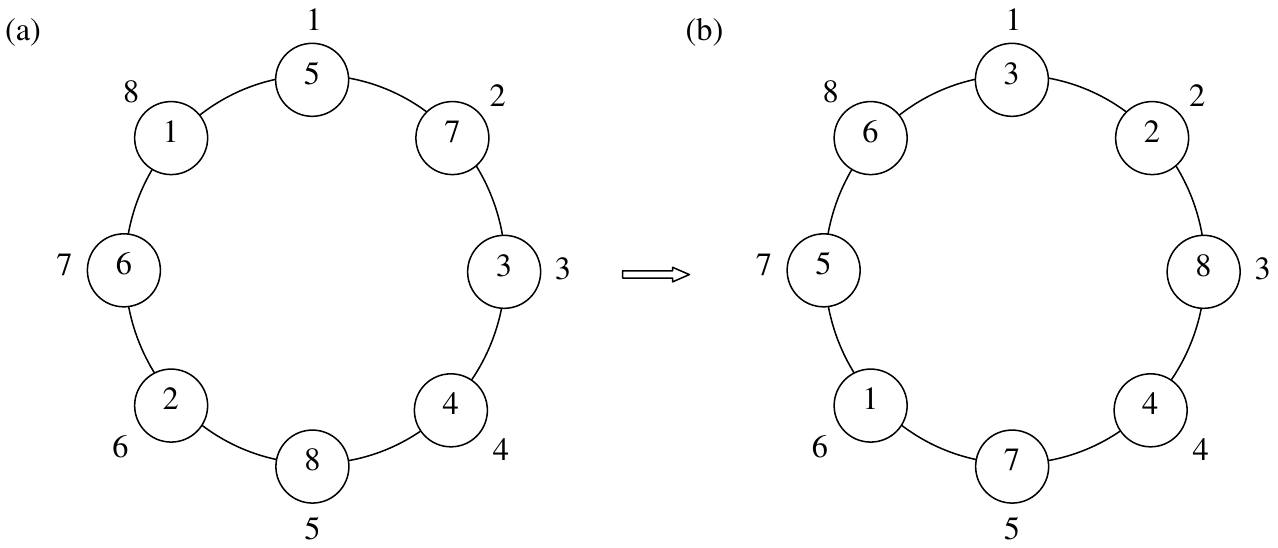}
\caption{Task swapping graphs of ring topology.}
\label{fig:RingTopology}
\end{figure}

A task swapping graph of ring topology is a circular task swapping graph in which each agent has direct links with exactly two other agents in the topology. In case any direct link of two agents is removed, a ring topology is changed into a line topology~\cite{Zhang2006}. We assume that a task swapping graph of ring topology having $n$ agents for $n \geq 3$ is labeled in such a way that $n$ agents are labeled clockwise in ascending order starting from 1 to $n$ (see Figure~\ref{fig:RingTopology}). Now, we denote a task swapping graph of ring topology having $n$ agents with their task assignment represented by permutation $p \in \mathfrak{S}_n$ as $\Gamma_p^{R(n)}$. We call the corresponding permutation $p \in \mathfrak{S}_n$ for $\Gamma_p^{R(n)}$ as a \emph{circular permutation}~\cite{Feng2011} in which position $i$ of a circular permutation $p \in \mathfrak{S}_n$ is referred to as verex (agent) $i$ for $1 \leq i \leq n$ 
 in $\Gamma_p^{R(n)}$. For example, a task swapping graph of Figure~\ref{fig:RingTopology}(a) is denoted as $\Gamma_{p_1}^{R(8)}$ and a task swapping graph of Figure~\ref{fig:RingTopology}(b) is denoted as $\Gamma_{p_2}^{R(8)}$, respectively,  where $p_1=5\,7\,3\,4\,8\,2\,6\,1 \in \mathfrak{S}_8$ and $p_2=3\,2\,8\,4\,7\,1\,5\,6 \in \mathfrak{S}_8$. 

Now, observe that a minimum-length sequence of adjacent task swappings needed from $\Gamma_{\pi_1}^{R(n)}$ to reach $\Gamma_{\pi_2}^{R(n)}$ is obtained by finding a minimum-length permutation factorization of $\pi_1^{-1}\pi_2$ using the generating set $S_6=\{(i\,i+1):1 \leq i < n\} \cup \{(1\,n)\}$. Note that adjacent task swappings in $\Gamma_{p}^{R(n)}$ are adjacent task swappings in $\Gamma_{p}^{L(n)}$ along with an adjacent task swapping between agent 1 and agent $n$. 

 A \emph{displacement vector}~\cite{Jerrum1985, Feng2011} $d=(d_1, d_2, d_3,\ldots, d_n)$ of a circular permutation $p \in \mathfrak{S}_n$ is introduced to sort a circular permutation $p$ into $I$ using the generating set $S_6=\{(i\,i+1):1 \leq i < n\} \cup \{(1\,n)\}$ by which we obtain a permutation factorization of $p$ using the same generating set. Each component $d_i$ in $d$ is defined as $d_i=j-i$, where $p(j)=i$ for $1 \leq i, j \leq n$. For any displacement vector $d=(d_1, d_2, d_3,\ldots, d_n)$, we have $\sum_{i=1}^n{d_i}=0$, and each $d_i=0$ if $d$ is a displacement vector of the identity permutation of $\mathfrak{S}_n$. For example, a displacement vector of a circular permutation $5\,7\,3\,4\,8\,2\,6\,1 \in \mathfrak{S}_8$ in Figure~\ref{fig:RingTopology}(a)  
is $(7, 4, 0, 0, -4, 1, -5, -3)$. Intuitively, each $|d_i|$ in $d$ is interpreted as the length of a path from position (vertex) $i$ to position (vertex) $k$ on which element $i$ is placed for a circular permutation $p \in \mathfrak{S}_n$, where $d_i$ is signed positive if the path from position $i$ to position $k$ is clockwise, and signed negative if the path from position $i$ to position $k$ is counterclockwise. We denote the corresponding path as $\text{path}(d_i)$, which is uniquely determined by $d_i$ of its circular permutation $p \in \mathfrak{S}_n$. Let $d_s$ be the maximum-valued component of a displacement vector $d$ of a circular permutation $p \in \mathfrak{S}_n$, and $d_t$ be the minimum-valued component of the displacement vector $d$. Since $\sum_{i=1}^n{d_i}=0$, $d_s$ is greater than 0 and $d_t$ is less than 0 for any non-identity circular permutation $p \in \mathfrak{S}_n$. If $d_s - d_t > n$ for each pair of indices $s$ and $t$, then we renew $d_s$ as $d_s-n$ and $d_t$ as $d_t+n$, respectively. This process is called \emph{strictly contracting transformation}~\cite{Feng2011}. If a displacment vector $d$ admits no strictly contracting transformation, we say that a displacement vector $d$ is \emph{stable}, denoted $\bar{d}$. For example, the maximum and minimum component values of displacement vector $d^\prime = (7, 4, 0, 0, -4, 1, -5, -3)$ of the circular permutation $5\,7\,3\,4\,8\,2\,6\,1 \in \mathfrak{S}_8$ in Figure~\ref{fig:RingTopology}(a) are $d_1^\prime=7$ and $d_7^\prime=-5$, respectively. Since $d_1^\prime - d_7^\prime =12 > 8$, we renew $d_1^\prime$ as $d_1^\prime=7-8=-1$, and $d_7^\prime=-5+8=3$. This procedure continues until we obtain a stable displacement vector $\bar{d}^\prime$, i.e., no pair of maximum-valued component $d_{s}^\prime$ and the minimum-valued component $d_t^\prime$ of $d^\prime$ satisfies $d_s^\prime-d_t^\prime > 8$. We leave it for the reader to verify that $\bar{d}^\prime = (-1, 4, 0, 0, -4, 1, 3, -3)$.

The value $\sum_{i=1}^n{|d_i|}$ of a stable displacement vector $\bar{d}$ of a circular permutation $p \in \mathfrak{S}_n$ is a key indicator of how close a circular permutation $p \in \mathfrak{S}_n$ is to the identity permutation in terms of a length using the generating set $S_6$. Note that $\sum_{i=1}^n{|d_i|}$ is not zero for any non-identity permutation in $\mathfrak{S}_n$, while $\sum_{i=1}^n{|d_i|}$ is 0 for the identity permutation. By using a stable displacement vector $\bar{d}$, an inversion number $I(\bar{d})$ is defined as $I(\bar{d})=|\{(i, j) : (i+\bar{d_i} > j+\bar{d_j}) \cup (i +\bar{d_i} + n < j  +\bar{d_j}),\;1 \leq i <j \leq n\}|$, which is the minimum length of a permutation factorization of $p \in \mathfrak{S}_n$ using the generating set  $S_6=\{(i\,i+1):1 \leq i < n\} \cup \{(1\,n)\}$~\cite{Jerrum1985}. Now, at each step of sorting a circular permutation $p \in \mathfrak{S}_n$ into $I$, we find an adjacent swapping to reduce an inversion number by 1. Observe that if an adjacent pair of positions (vertices) $v_1$ and $v_2$ have elements $s$ and $t$, respectively, such that $\text{path}(s)$ and $\text{path}(t)$ are directed oppositely having an intersection of edge $(v_1, v_2)$, then swapping elements $s$ and $t$ on vertices $v_1$ and $v_2$ reduces an inversion number by 1. Observe also the case where an adjacent pair of vertices $v_1$ and $v_2$ have elements $s$ and $t$, respectively, such that $s$ is homed (i.e., $p(s)=s$) and $t$ is not homed. If $\text{path}(t)$ crosses vertex $v_1$, then swapping elements $s$ and $t$ on vertices $v_1$ and $v_2$ reduces an inversion number by 1. Now, each step of the sorting procedure is to find an adjacent swapping that reduces an inversion number by 1. As stated earlier in this section, if $\bar{d}$ is a stable displacement vector of a circular permutation $p \in \mathfrak{S}_n$, then the inversion number $I(\bar{d})$ is the minimum length of a permutation factorization of $p$ using the generating set $S_6$. Therefore, $I(\bar{d})$ is the minimum number of adjacent swappings required for sorting a circular permutation $p \in \mathfrak{S}_n$ to the identity permutation $I$ using the generating set $S_6$.

\begin{algorithm}[h!]
\SetAlgoLined
\KwIn{A source and a target task assignment in a task swapping graph of ring topology $\Gamma_{\pi_1}^{R(n)}$ and $\Gamma_{\pi_2}^{R(n)}$, respectively.}
\KwOut{A minimum-length sequence of adjacent task swappings needed from $\Gamma_{\pi_1}^{R(n)}$ to reach $\Gamma_{\pi_2}^{R(n)}$.}
\tcp*[h]{Find a minimum-length permutation factorization of $\pi_1^{-1}\pi_2$ by sorting $\pi_2^{-1}\pi_1$ to the identity permutation $I$ using the generating set $S_6=\{(i\,i+1):1 \leq i < n\} \cup \{(1\,n)\}$}\break
\Begin
{
Let $\pi=\pi_2^{-1}\pi_1 \in \mathfrak{S}_n$. Find a stable displacement vector $\bar{d}$ of $\pi$, and calculate the inversion number $I(\bar{d})$\;
\While{$\pi \neq I$}
{
In a circular permutation $\pi \in \mathfrak{S}_n$ in $\Gamma_{\pi}^{R(n)}$, find an adjacent pair of vertices $v_1$ and $v_2$ having elements $s$ and $t$, respectively, such that $\text{path}(s)$ and $\text{path}(t)$ derived from its stable displacement vector of $\pi$ are directed oppositely having an intersection of edge $(v_1, v_2)$. If no such an adjacent pair exists, then find an adjacent pair of vertices $v_1$ and $v_2$ having elements $s$ and $t$, respectively, such that $s$ is homed (i.e., $\pi(s)=s$) and $t$ is not homed in which $\text{path}(t)$ crosses vertex $v_1$\;

Swap elements $s$ and $t$ on vertices $v_1$ and $v_2$, replacing $\pi$ with the resulting permutation. Compute a stable displacement vector of the (updated) circular permutation $\pi$\; 
}

Obtain a minimum-length sequence of adjacent task swappings needed from $\Gamma_{\pi_1}^{R(n)}$ to reach $\Gamma_{\pi_2}^{R(n)}$ by using the above permutation factorization of $\pi_1^{-1}\pi_2$\;
}
\caption{A task reassignment by using adjacent task swappings in a task swapping graph of ring topology.}
\label{Algorithm:TSGRing}
\end{algorithm}

Algorithm~\ref{Algorithm:TSGRing} describes the procedure of converting $\Gamma_{\pi_1}^{R(n)}$ into $\Gamma_{\pi_2}^{R(n)}$ by using a minimum-length sequence of adjacent task swappings. By using Algorithm~\ref{Algorithm:TSGRing}, we obtain a minimum-length sequence of adjacent task swappings needed from a task swapping graph $\Gamma_{p_1}^{R(8)}$ in Figure~\ref{fig:RingTopology}(a) to reach a task swapping graph $\Gamma_{p_2}^{R(8)}$ in Figure~\ref{fig:RingTopology}(b), where $p_1=5\,7\,3\,4\,8\,2\,6\,1 \in \mathfrak{S}_8$ and $p_2=3\,2\,8\,4\,7\,1\,5\,6 \in \mathfrak{S}_8$. A direct computation shows that $p_2^{-1}p_1$ is $7\,5\,1\,4\,3\,2\,8\,6$ and its stable displacement vector is $(2, -4, 2, 0, -3, 2, 2, -1)$. An adjacent swapping between 7th position (element 8) and 8th position (element 6) of $p_2^{-1}p_1$ reduces an inversion number by 1. This sorting procedure is continued by using Algorithm~\ref{Algorithm:TSGRing} until arriving at the identity permutation (see below):\\
$7\,5\,1\,4\,3\,2\,8\,6 \xrightarrow{<7, 8>} 7\,5\,1\,4\,3\,2\,6\,8$$ \xrightarrow{<2, 3>} 7\,1\,5\,4\,3\,2\,6\,8 \xrightarrow{<6, 7>} 7\,1\,5\,4\,3\,6\,2\,8  $$\xrightarrow{<3, 4>}\\  7\,1\,4\,5\,3\,6\,2\,8   \xrightarrow{<4, 5>} 7\,1\,4\,3\,5\,6\,2\,8  $$\xrightarrow{<7, 8>} 7\,1\,4\,3\,5\,6\,8\,2  \xrightarrow{<1, 8>} 2\,1\,4\,3\,5\,6\,8\,7   $$\xrightarrow{<1, 2>}\\ 1\,2\,4\,3\,5\,6\,8\,7  \xrightarrow{<7, 8>} 1\,2\,4\,3\,5\,6\,7\,8 $$ \xrightarrow{<3, 4>} 1\,2\,3\,4\,5\,6\,7\,8$.

Now, we have the resulting minimum-length sequence of adjacent task swappings needed from $\Gamma_{p_1}^{R(8)}$ to reach $\Gamma_{p_2}^{R(8)}$, which is $((7\,8)$, $(2\,3)$, $(6\,7)$, $(3\,4)$, $(4\,5)$, $(7\,8)$, $(1\,8)$, $(1\,2)$, $(7\,8)$, $(3\,4))$ of length 10. 

Note that an upper bound of the number of adjacent task swappings needed from $\Gamma_{\pi_1}^{R(n)}$ to reach $\Gamma_{\pi_2}^{R(n)}$ is subject to the diameter of the modified bubble-sort graph $MBS_n$ discussed in Section~\ref{sec:TranspositionGraphs}. To the best of our knowledge, the formula of the diameter of $MBS_n$ is not known~\cite{Feng2011, Lakshmivarahan1993, Stacho1998}. Nevertheless, a minimum-length sequence of sorting an aribtrary circular permutation $p \in \mathfrak{S}_n$ into the identity permutation $I$ using the generating set $S_6=\{(i\,i+1):1 \leq i < n\} \cup \{(1\,n)\}$ in Algorithm~\ref{Algorithm:TSGRing} is obtained in polynomial time~\cite{Jerrum1985}. It follows that Algorithm~\ref{Algorithm:TSGRing} runs in polynomial time as with other algorithms involving permutation sortings discussed in this paper.

\begin{algorithm}[h!]
\SetAlgoLined
\KwIn{A source and a target task assignment in a task swapping graph of an arbitrary tree topology $\Gamma_{\pi_1}^{T(n)}$ and $\Gamma_{\pi_2}^{T(n)}$, respectively.}
\KwOut{A sequence of adjacent task swappings needed from $\Gamma_{\pi_1}^{T(n)}$ to reach $\Gamma_{\pi_2}^{T(n)}$ in the number $c(\pi) -n + \sum_{i=1}^n{d(i, \pi(i))}$ of steps, where $\pi=\pi_2^{-1}\pi_1$, $c(\pi)$ is the number of cycles in $\pi$, and $d(i, j)$ is the distance between agent $i$ and agent $j$ in $\Gamma_{\pi}^{T(n)}$.}

\Begin
{
Let $\pi=\pi_2^{-1}\pi_1$ and start the procedure of sorting $\Gamma_{\pi}^{T(n)}$ to $\Gamma_{I}^{T(n)}$\;

\While{$\pi \neq I$}
{
In $\Gamma_{\pi}^{T(n)}$ find an adjacent pair of agents $a_1$ and $a_2$ such that their unhomed tasks $t_1$ and $t_2$, respectively, need to move toward each other for their homed positions (i.e., $\pi(t_k) = t_k$ for $k=1$ and $k= 2$). Or find an adjacent pair of agents $a_1$ and $a_2$ such that its task $t_1$ is homed and its task $t_2$ is not homed, respectively, in which task $t_2$ needs to move toward and cross agent $a_1$ for its homed position (i.e., $\pi(t_2) = t_2$). Then, swap task $t_1$ on agent $a_1$ and task $t_2$ on agent $a_2$, replacing $\pi$ with the resulting permutation\;
}
Obtain a sequence of adjacent task swappings needed from $\Gamma_{\pi_1}^{T(n)}$ to reach $\Gamma_{\pi_2}^{T(n)}$ by using the above sorting procedure\;
}
\caption{A task reassignment by using adjacent task swappings in a task swapping graph of an arbitrary tree topology.}
\label{Algorithm:TSGTree}
\end{algorithm}

Finally, we discuss a task swapping graph of an arbitrary tree topology that has not been discussed in this section. A task swapping graph of a tree topology having $n$ agents with their task assignment represented by permutation $p \in \mathfrak{S}_n$ is denoted as $\Gamma_p^{T(n)}$. We concern the procedure of converting a source task assignment in $\Gamma_{\pi_1}^{T(n)}$ to a target task assignment in $\Gamma_{\pi_2}^{T(n)}$ by using the minimum number of adjacent task swappings. 

Due to Corollary~\ref{cor:diameterOfCayleygraph}, we may not always obtain a tight upper bound for the number of steps to convert $\Gamma_{\pi_1}^{T(n)}$ into $\Gamma_{\pi_2}^{T(n)}$ for $\pi_1 \in \mathfrak{S}_n$ and $\pi_2 \in \mathfrak{S}_n$. However, we may apply Theorem~\ref{theorem:TranspositionTree} to convert $\Gamma_{\pi_1}^{T(n)}$ into $\Gamma_{\pi_2}^{T(n)}$ in the number $c(\pi) -n + \sum_{i=1}^n{d(i, \pi(i))}$ of steps, where $\pi=\pi_2^{-1}\pi_1$, $c(\pi)$ is the number of cycles in $\pi$, and $d(i, j)$ is the distance between position (vertex) $i$ and position (vertex) $j$ in $\Gamma_{\pi}^{T(n)}$. Algorithm~\ref{Algorithm:TSGTree} describes the procedure of converting $\Gamma_{\pi_1}^{T(n)}$ into $\Gamma_{\pi_2}^{T(n)}$ in the above number of steps.

\begin{proposition}
An upper bound for the number of adjacent task swappings needed from $\Gamma_{\pi_1}^{T(n)}$ for $\pi_1 \in \mathfrak{S}_n$ to reach $\Gamma_{\pi_2}^{T(n)}$ for $\pi_2 \in \mathfrak{S}_n$ is $c(\pi) -n + \sum_{i=1}^n{d(i, \pi(i))}$, where $\pi = \pi_2^{-1}\pi_1$ and $c(\pi)$ is the number of cycles in $\pi$.
\end{proposition}
\begin{proof}
It follows directly from Theorem~\ref{theorem:TranspositionTree}, Corollary~\ref{cor:diameterOfCayleygraph}, and Algorithm~\ref{Algorithm:TSGTree} in which finding a sequence of legal moves of a permutation puzzle from the given initial position corresponding to permutation $\pi_2^{-1}\pi_1$ for $\pi_1, \pi_2 \in \mathfrak{S}_n$ to the final position corresponding to permutation $I$ using the procedure discussed at the end of Section~\ref{sec:TranspositionGraphs} is converted to the context of finding a sequence of adjacent task swappings needed from $\Gamma_{\pi_1}^{T(n)}$ to reach $\Gamma_{\pi_2}^{T(n)}$. $\qquad$
\end{proof}

\section{Related Work and Implementation}
\label{sec:RelatedWork}
The linear assignment problem and its variants are one of the fundamental problems in both computer science and operations research~\cite{Bowen1992, Burkard2009, Zavlanos2007, Zavlanos2008}. However, little research has been done for bijective task reassignments using iterative local (adjacent) task swappings among agents in a network topology. From the known results of Cayley graphs and permutation factorizations, we have applied them to our task swapping networks of several well-known topologies. Cayley graph approaches to interconnection networks have been researched in~\cite{Akers1989, Heydemann1997, Lakshmivarahan1993, Zhu1996}, but no (bijective) task assignment of tasks to agents is involved in them. Similarly, the group-theoretic approaches for bijective task assignments have been discussed in~\cite{Gutjahr1997, Kim2010, Rowe2002}, but task swappings among agents in a network topology have not been considered. Although minimum-generator sequence and cycle factorization problems have been researched in~\cite{Irving2009, Jerrum1985, Mackiw1995, Pak1999, Zhu1996}, no task swapping or task assignment is considered in them. Meanwhile, permutation puzzles on transposition trees have been briefly discussed in~\cite{Akers1989}. However, transposition trees in~\cite{Akers1989} simply denote transpositions along with generating sets for permutation groups, which do not intend to represent task assignments involving network topologies. By assigning IDs of $n$ agents from 1 to $n$ in a predetermined manner corresponding to a given network topology, task reassignments of $n$ tasks using iterative local task swappings are represented by purely algebraic forms in our approach.

We have developed an essential execution environment\footnote{Source codes and sample data are available at http://www.airesearch.kr/downloads/tsg.zip} of our approach using GNU C++~\cite{GNU}. We have verified Algorithm~\ref{Algorithm:TSGLine}$\sim$\ref{Algorithm:TSGRing} in this paper using our implementation up to 200 simulated tasks and agents. As briefly discussed in Section~\ref{sec:TaskSwappingNetworks}, Algorithm~\ref{Algorithm:TSGLine}$\sim$\ref{Algorithm:TSGRing} in this paper run in polynomial time. Our implementation is generic, in the sense that tasks in task swapping networks can be interchangeable with tokens or objects. Our implementation shows that it can further be employed for a subclass of object or token sorting problems using adjacent swappings in a given well-known network topology discussed in this paper.

\section{Conclusions}
This paper presented task swapping networks of several basic topologies used in distributed systems. Task swappings between adjacent agents in a network topology are represented by task swappings of swapping distance 1 in the corresponding task swapping graph. We considered the situation in which the total cost of task migrations relies on the number of adjacent task swappings involved in a given network topology. Minimum generator sequence algorithms using several known generating sets for $\mathfrak{S}_n$ allow us to find a minimum-length sequence of adjacent task swappings needed from a source task assignment to reach a target task assignment in a task swapping graph of several topologies, such as line, star, complete, complete bipartite, and ring. 

Task swapping graphs of the more complex topologies (e.g., 2D and 3D grids, hypercubes, etc) along with task swappings of swapping distance $k \geq 2$ have not been discussed in this paper. It is a challenging research problem to find whether or not there exists a polynomial-time algorithm of finding a minimum-length sequence of adjacent task swappings needed from a source task assignment to reach a target task assignment in a task swapping graph of 2D (respectively, 3D) grid topology. We leave it as an open problem. 

\label{sec:Conclusion}

\bibliographystyle{gCOM}
\bibliography{dkim}

\end{document}